\documentclass[letterpaper,english,10pt,aps,letter,superscriptaddress,nofootinbib,pre]{revtex4}
\pdfoutput=1
\usepackage[T1]{fontenc}
\usepackage[latin9]{inputenc}
\usepackage{xcolor}
\usepackage{pdfcolmk}
\usepackage{verbatim}
\usepackage{float}
\usepackage{amsmath}
\usepackage{amssymb}
\usepackage{graphicx}
\usepackage{esint}
\PassOptionsToPackage{normalem}{ulem}
\usepackage{ulem}

\makeatletter

\pdfpageheight\paperheight
\pdfpagewidth\paperwidth

\floatstyle{ruled}
\newfloat{algorithm}{tbp}{loa}
\providecommand{\algorithmname}{Algorithm}
\floatname{algorithm}{\protect\algorithmname}
\providecolor{lyxadded}{rgb}{1,0,0}
\providecolor{lyxdeleted}{rgb}{0,0,1}

\@ifundefined{textcolor}{}
{%
 \definecolor{BLACK}{gray}{0}
 \definecolor{WHITE}{gray}{1}
 \definecolor{RED}{rgb}{1,0,0}
 \definecolor{GREEN}{rgb}{0,1,0}
 \definecolor{BLUE}{rgb}{0,0,1}
 \definecolor{CYAN}{cmyk}{1,0,0,0}
 \definecolor{MAGENTA}{cmyk}{0,1,0,0}
 \definecolor{YELLOW}{cmyk}{0,0,1,0}
}

\usepackage{ae,aecompl}

\newcommand{\sfrac}[2]{\mathchoice
  {\kern0em\raise.5ex\hbox{\the\scriptfont0 #1}\kern-.15em/
   \kern-.15em\lower.25ex\hbox{\the\scriptfont0 #2}}
  {\kern0em\raise.5ex\hbox{\the\scriptfont0 #1}\kern-.15em/
   \kern-.15em\lower.25ex\hbox{\the\scriptfont0 #2}}
  {\kern0em\raise.5ex\hbox{\the\scriptscriptfont0 #1}\kern-.2em/
   \kern-.15em\lower.25ex\hbox{\the\scriptscriptfont0 #2}}
  {#1\!/#2}}

\makeatother

\usepackage{babel}
\begin{document}
\global\long\def\V#1{\boldsymbol{#1}}
\global\long\def\M#1{\boldsymbol{#1}}
\global\long\def\Set#1{\mathbb{#1}}

\global\long\def\D#1{\Delta#1}
\global\long\def\d#1{\delta#1}

\global\long\def\norm#1{\left\Vert #1\right\Vert }
\global\long\def\abs#1{\left|#1\right|}

\global\long\def\grad{\M{\nabla}}
\global\long\def\avv#1{\langle#1\rangle}
\global\long\def\av#1{\left\langle #1\right\rangle }

\global\long\def\Mob{\M M}
\global\long\def\J{\M B}
\global\long\def\S{\M B^{\star}}
\global\long\def\L{\M C}

\global\long\def\shalf{\sfrac{1}{2}}
\global\long\def\sthreehalves{\sfrac{3}{2}}

\global\long\def\myhalf{\frac{1}{2}}
\global\long\def\mythreehalves{\frac{3}{2}}

\title{Multiscale Temporal Integrators for Fluctuating Hydrodynamics}

\author{Steven Delong}

\affiliation{Courant Institute of Mathematical Sciences, New York University,
New York, NY 10012}

\author{Yifei Sun}

\affiliation{Courant Institute of Mathematical Sciences, New York University,
New York, NY 10012}

\author{Boyce E. Griffith}

\affiliation{Department of Mathematics, University of North Carolina, Chapel Hill,
NC 27599-3250}

\affiliation{Courant Institute of Mathematical Sciences, New York University,
New York, NY 10012}

\author{Eric Vanden-Eijnden}

\email{eve2@courant.nyu.edu}

\selectlanguage{english}%

\affiliation{Courant Institute of Mathematical Sciences, New York University,
New York, NY 10012}

\author{Aleksandar Donev}

\email{donev@courant.nyu.edu}

\selectlanguage{english}%

\affiliation{Courant Institute of Mathematical Sciences, New York University,
New York, NY 10012}
\begin{abstract}
Following on our previous work {[}S. Delong and B. E. Griffith and
E. Vanden-Eijnden and A. Donev, Phys. Rev. E, 87(3):033302, 2013{]},
we develop temporal integrators for solving Langevin stochastic differential
equations that arise in fluctuating hydrodynamics. Our simple predictor-corrector
schemes add fluctuations to standard second-order deterministic solvers
in a way that maintains second-order weak accuracy for linearized
fluctuating hydrodynamics. We construct a general class of schemes
and recommend two specific schemes: an explicit midpoint method, and
an implicit trapezoidal method. We also construct predictor-corrector
methods for integrating the overdamped limit of systems of equations
with a fast and slow variable in the limit of infinite separation
of the fast and slow timescales. We propose using random finite differences
to approximate some of the stochastic drift terms that arise because
of the kinetic multiplicative noise in the limiting dynamics. We illustrate
our integrators on two applications involving the development of giant
nonequilibrium concentration fluctuations in diffusively-mixing fluids.
We first study the development of giant fluctuations in recent experiments
performed in microgravity using an overdamped integrator. We then
include the effects of gravity, and find that we also need to include
the effects of fluid inertia, which affects the dynamics of the concentration
fluctuations greatly at small wavenumbers.
\end{abstract}
\maketitle

\section{Introduction and Background}

Fluctuating Hydrodynamics (FHD) accounts for stochastic effects arising
at mesoscopic and macroscopic scales because of the discrete nature
of fluids at microscopic scales \cite{Landau:Fluid,LLNS_FD_Fox,FluctHydroNonEq_Book}.
In FHD, spatially-extended Langevin equations are constructed by including
stochastic flux terms in the classical Navier-Stokes-Fourier equations
of fluid dynamics and related conservation laws. It is widely appreciated
that thermal fluctuations are important in flows at micro and nano
scales; even more importantly, hydrodynamic fluctuations span the
whole range of scales from the microscopic to the macroscopic and
need to be consistently included in \emph{all} levels of description
\cite{GiantFluctuations_Nature,DropletSpreading,FractalDiffusion_Microgravity,DiffusionJSTAT}.
While the original formulation of fluctuating hydrodynamics was for
compressible single-component fluids \cite{Landau:Fluid}, the methodology
can be extended to other systems such as fluid mixtures \cite{FluctHydroNonEq_Book},
chemically reactive systems \cite{FluctuatingReactionDiffusion},
magnetic materials \cite{FluctHydro_Magnetic}, and others \cite{LebowitzHydroReview}.
The structure of the equations of fluctuating hydrodynamics can be,
to some extent, justified on the basis of the Mori-Zwanzig formalism
\cite{LLNS_Espanol,DiscreteLLNS_Espanol}. The basic idea is to add
a \emph{stochastic flux} corresponding to each dissipative (irreversible,
diffusive) flux, leading to a continuum Langevin model that ensures
detailed balance with respect to a suitable Einstein or Gibbs-Boltzmann
equilibrium distribution \cite{OttingerBook}.

After spatial discretization (truncation) of the stochastic partial
differential equations (SPDEs) of FHD, one obtains a large-scale system
of stochastic ordinary differential equations (SODEs) that has the
familiar structure of Langevin equations common in statistical mechanics.
The spatial discretization must be performed with specific attention
to preserving fluctuation-dissipation balance \cite{DFDB}; alternatively,
one can directly construct discrete Langevin equations with the proper
structure from the underlying microscopic dynamics by using the theory
of coarse-graining \cite{DiscreteDiffusion_Espanol,DiscreteLLNS_Espanol}.
In this paper, continuing on our previous work \cite{DFDB}, we are
concerned with temporal integration of the Langevin systems that arise
in fluctuating hydrodynamics and other mesoscopic models. In \cite{DFDB},
we constructed several simple predictor-corrector (two-step) schemes
for Langevin equations and applied them to the incompressible fluctuating
hydrodynamics equations coupled with an advection-diffusion equation
for a scalar concentration field. The main feature of these schemes
is that they allow one to treat some terms implicitly (e.g., mass
or momentum diffusion), and treat others explicitly (e.g, advection).
Furthermore, the schemes presented in \cite{DFDB} are second-order
weakly accurate for nonlinear SODEs with \emph{constant additive noise},
while also achieving, in certain cases, third-order accuracy for the
static correlation functions (static structure factors in fluctuating
hydrodynamics).

In practical applications, however, several difficulties arise that
require the development of novel temporal integration schemes. The
first complication in FHD is the appearance of multiplicative noise.
In the context of SPDEs, such multiplicative noise is often purely
formal and cannot be interpreted mathematically in continuum formulations
unless the nonlinear terms are regularized \cite{DiffusionJSTAT,DDFT_Hydro},
or suitable renormalization terms are added \cite{RegularityStructures}.
In most cases, a precise formulation of the multiplicative noise terms
is not known and the importance of various stochastic drift terms
arising due to the multiplicative nature of the noise have not been
explored. A precise mathematical interpretation can, however, be given
to the equations of \emph{linearized} FHD (LFHD), which are in fact
the most common model used in the literature \cite{FluctHydroNonEq_Book}.
The LFHD equations can in some cases be derived rigorously from the
microscopic dynamics as a form of central limit theorem for the Gaussian
fluctuations around the deterministic hydrodynamic equations, which
are themselves a form of law of large numbers for the macroscopic
observables \cite{LebowitzHydroReview,MicroToSPDE_Review,AERW_Varadhan,LDT_Excluded,LLN_InteractingBrownian}.
One can, at least formally, obtain the LFHD equations from the nonlinear
FHD equations by expanding to leading order in the magnitude of the
stochastic forcing terms (more precisely, the inverse of the coarse-graining
length scale).

Such linearization of nonlinear FHD equations leads to a system of
two equations, the usual \emph{nonlinear} deterministic equation,
and a \emph{linear} (additive-noise) stochastic differential equation
(SDE) for the Gaussian fluctuations around the mean. A naive application
of the temporal integrators in \cite{DFDB} would require first solving
the deterministic equation, itself a nontrivial problem except in
the simplest of cases, and \emph{then} solving a linear SDE with a
time-dependent but additive noise of magnitude determined by the deterministic
solution. Here we demonstrate that with some simple modifications
the predictor-corrector integrators from \cite{DFDB} can accomplish
these two steps together without ever explicitly writing the LFHD
equations. Specifically, here we construct schemes that numerically
linearize the equations around a numerically-determined deterministic
solution. Our analysis shows how to achieve second-order weak accuracy
for the LFHD equations by choosing where to evaluate the amplitude
of the stochastic forcing terms. In certain cases, the resulting integrators
will also be first-order weakly accurate for the (discrete or regularized)
\emph{nonlinear} FHD equations. For nonlinear Langevin equations with
multiplicative noise, in general, it is difficult to construct integrators
of weak order higher than one. Second-order weak Runge-Kutta (derivative-free)
schemes have been constructed \cite{WeakSecondOrder_RK} for systems
of SODEs, however, using these types of methods in the contexts of
SPDEs is non trivial.

A second difficulty that often arises in FHD is the appearance of
large separation of time scales between the different hydrodynamic
variables. In incompressible FHD, velocity fluctuations are the most
rapid, since flows at small scales are typically viscous-dominated
and momentum diffuses much faster than does mass. As the dynamics
of interest is usually the slow dynamics of the concentration field
or discrete particles advected by the velocity fluctuations, specialized
multiscale temporal integrators are required to avoid the need to
use small time step sizes that resolve the fast dynamics. This can
be accomplished by analytically performing adiabatic mode elimination
and eliminating the fast variable from the description to obtain a
limiting or \emph{overdamped} equation for the slow variables, and
then numerically integrating the limiting dynamics. In this work we
develop predictor-corrector schemes that in essence numerically take
the overdamped limit of a two-scale (fast-slow) system of SODEs in
which the fast variable enters linearly. Our predictor-corrector schemes
are applicable to a broad range of two-scale Langevin equations that
frequently arise in practice in a variety of contexts. Their key feature
is that they obtain all of the stochastic drift terms numerically
without requiring derivatives. This makes it relatively easy to take
a code that integrates the original fast-slow \emph{inertial} dynamics
using a (semi-)implicit method, and to convert it into a code that
integrates the overdamped dynamics. Furthermore, the schemes we construct
here are second-order weakly accurate for the linearized overdamped
equations. In order to facilitate the integration of our methods in
existing codes our algorithms make use of the components already required
to integrate the original fast-slow equations without making use of
the large-separation of scales. The result is a new algorithm that
reuses the base code but can take a time step several orders of magnitude
larger.

In this work we demonstrate that simple predictor-corrector methods
can address the difficulties discussed above with a minimal amount
of effort on the part of the user. Specifically, we develop simple
predictor-corrector schemes for temporal integration of Langevin SDEs
that arise in fluctuating hydrodynamics and accomplish the following
design goals:
\begin{enumerate}
\item They reuse the same computational components already available in
standard computational fluid dynamics (CFD) solvers, such as linear
solvers for viscosity or diffusion, advection schemes, etc.
\item In the deterministic context they are second-order accurate and relatively
standard.
\item For nonlinear but additive noise SDEs they are second-order weakly
accurate \cite{DFDB}.
\item They numerically linearize the FHD equations and solve the LFHD equations
weakly to second-order.
\item For LFHD, they give higher-order accurate static correlations (static
structure factors) at steady state \cite{DFDB}.
\item They can be used to integrate two-scale fast-slow systems in the overdamped
limit, preserving all of the above properties but now referring to
the limiting dynamics.
\item In the general nonlinear multiplicative-noise case, they are weakly
first-order accurate and do not require any derivative information.
\end{enumerate}
Some of us have previously made use of some of the temporal integrators
described here in more specialized works. In \cite{BrownianBlobs}
we constructed temporal integrators for the equations of Brownian
Dynamics, which is a multiplicative-noise overdamped SDE. These overdamped
Langevin equations result when one eliminates the fast velocity degrees
of freedom from a system of equations for the motion of particles
immersed in a fluctuating Stokes fluid. In \cite{DiffusionJSTAT}
we solved a related overdamped SPDE for the concentration of the immersed
particles \cite{DDFT_Hydro} using the techniques detailed here. In
\cite{LowMachImplicit} we apply our methods to the low Mach number
equations of FHD. In this work we present the temporal integrators
used in this prior work in greater generality, with the hope that
they will be useful for other Langevin equations. We also explain
the analysis required to study the order of weak accuracy of the schemes,
making it easier for other researchers to further generalize our methods
or to apply them in specific contexts.

The remainder of this section discusses Langevin equations and a specific
example thereof: the equations of fluctuating hydrodynamics. In section
\ref{sec:Linearized} we introduce schemes that are second order weakly
accurate for the linearized (weak noise limit) Langevin equations.
Section \ref{sec:Overdamped} studies systems of equations with a
fast and slow variable and proposes schemes which perform adiabatic
elimination of the fast variable numerically. Section \ref{sec:Applications}
outlines specific versions of our schemes applied to a system of passive
tracers advected by a fluctuating fluid, and in section \ref{sec:GiantFluct},
we use our temporal integration methods to simulate diffusive mixing
in binary liquid mixtures. Finally, section \ref{sec:Conclusions}
gives our concluding thoughts and remarks.

\subsection{\label{sec:GLE}Generic Langevin equations}

As discussed in more detail in our previous work \cite{DFDB}, we
consider temporal integrators for a class of generic Langevin equations
for the coarse-grained variables $\V x\left(t\right)$ \cite{GrabertBook},
\begin{equation}
\partial_{t}\V x=-\M N\left(\V x\right)\frac{\partial U}{\partial\V x}+\left(2k_{B}T\right)^{1/2}\M M_{\myhalf}\left(\V x\right)\,\M{\mathcal{W}}(t)+\left(k_{B}T\right)\frac{\partial}{\partial\V x}\cdot\M N\left(\V x\right),\label{eq:x_t_general}
\end{equation}
where $\M{\mathcal{W}}(t)$ denotes white noise, the formal time derivative
of a collection of independent Brownian motions, and an Ito interpretation
is assumed. Here $U\left(\V x\right)$ is the thermodynamic driving
potential, such as an externally-applied conservative potential or
a \emph{coarse-grained free energy.} We will typically suppress the
explicit dependence on $\V x$ and write the \emph{mobility operator}
as $\M N\equiv\M N\left(\V x\right)$, and assume that the noise operator
$\M M_{\myhalf}\left(\V x\right)$ obeys the \emph{fluctuation-dissipation
balance} condition
\begin{equation}
\M M_{\myhalf}\M M_{\myhalf}^{\star}=\M M=\frac{1}{2}\left(\M N+\M N^{\star}\right),\label{eq:FDB}
\end{equation}
where star denotes an adjoint. This ensures that the dynamics (\ref{eq:x_t_general})
is (under suitable assumptions) ergodic with respect to the Gibbs-Boltzmann
distribution
\begin{equation}
P_{\text{eq}}\left(\V x\right)=Z^{-1}\,\exp\left[-\frac{U\left(\V x\right)}{k_{B}T}\right],\label{eq:GibbsBoltzmann}
\end{equation}
where $Z$ is a normalization constant. In the notation we have chosen,
in component form $\left(\partial_{\V x}\cdot\M N\right)_{i}=\partial_{j}N_{ij}$,
where repeated indices are implicitly summed throughout this paper.
Please note that this is different from the less-standard notation
we used in our previous paper \cite{DFDB}, where the drift term is
written as $\left(k_{B}T\right)\,\partial_{\V x}\cdot\M N^{\star}$.

The last term in (\ref{eq:x_t_general}) is an additional ``stochastic''
or ``thermal'' drift term, as can be seen from the fact that it
is proportional to $k_{B}T$. This term contains information about
the noise and therefore depends on the particular interpretation of
the stochastic integral. The thermal drift term, however, also contains
information about the anti-symmetric part of the generator $\M N$
that is unrelated to the stochastic noise term. If the skew-adjoint
component $\M L=\frac{1}{2}\left(\M N^{\star}-\M N\right)$ is divergence-free,
$\partial_{\V x}\cdot\M L=0$, the associated dynamics is incompressible
in phase space \cite{AugmentedLangevin} and the stochastic drift
term disappears if one uses the kinetic interpretation \cite{KineticStochasticIntegral_Ottinger}
of the stochastic integral, denoted with the stochastic product $\diamond$,
\begin{equation}
\partial_{t}\V x=-\M N\left(\V x\right)\frac{\partial U}{\partial\V x}+\left(2k_{B}T\right)^{1/2}\M M_{\myhalf}\left(\V x\right)\diamond\M{\mathcal{W}}(t).\label{eq:x_t_kinetic}
\end{equation}
In the case when $\M N=\M M$ is self-adjoint, $\M L=0$, the kinetic
SDE (\ref{eq:x_t_kinetic}) corresponds to a Fokker-Planck equation
for the evolution of the probability distribution $P\left(\V x,t\right)$
for observing the state $\V x$ at time $t$, 
\begin{equation}
\frac{\partial P}{\partial t}=\frac{\partial}{\partial\V x}\cdot\left\{ \Mob\left[\frac{\partial U}{\partial\V x}P+\left(k_{B}T\right)\frac{\partial P}{\partial\V x}\right]\right\} ,\label{eq:FokkerPlanck}
\end{equation}
where we note that the term in square brackets vanishes when $P=P_{eq}$.

\subsection{\label{sub:OverdampedLang}Overdamped Langevin equation}

As a simple but illustrative example of a Langevin equation that our
methods can be applied to we consider the Langevin equation with position-dependent
friction tensor $\M{\gamma}\left(\V x\right)$,
\begin{align}
m\partial_{t}\V v & =\V F\left(\V x\right)-\M{\gamma}\left(\V x\right)\V v+\sqrt{2k_{B}T\M{\gamma}\left(\V x\right)}\,\V{\mathcal{W}}\left(t\right)\nonumber \\
\partial_{t}\V x & =\V v,\label{eq:inertial_Langevin}
\end{align}
where $\V x(t)$ is the position of a particle diffusing under the
influence of an external force $\M F(\V x)=-\partial_{\V x}U(\V x)$,
$\V v(t)$ is the particle velocity, and $m$ is its mass. It is well-known
that in the low-inertia limit $m\rightarrow0$, the velocity can be
eliminated as a fast degree of freedom to obtain the overdamped Langevin
equation \cite{ModeElimination_Papanicolaou,AveragingHomogenization}
\begin{equation}
\partial_{t}\V x=\M{\gamma}^{-1}\left(\V x\right)\V F\left(\V x\right)+\sqrt{2k_{B}T}\,\M{\gamma}^{-\frac{1}{2}}\left(\V x\right)\V{\mathcal{W}}\left(t\right)+\left(k_{B}T\right)\partial_{\V x}\cdot\M{\gamma}^{-1}\left(\V x\right),\label{eq:overdamped_Langevin}
\end{equation}
which is of the form (\ref{eq:x_t_general}) with a self-adjoint mobility
$\M N=\M M\equiv\M{\gamma}^{-1}$ and with $\M M^{\frac{1}{2}}\equiv\M{\gamma}^{-\frac{1}{2}}$
being a matrix square root or a Cholesky factor of $\M{\gamma}^{-1}$.
Even though the overdamped equation strictly only applies in the limit
of infinite separation of time scales, it will be a very good approximation
so long as there still remains a large separation of time scales between
the fast velocity and the slow position.

\subsubsection{\label{sub:Fixman}Fixman Scheme}

In this paper we construct predictor-corrector algorithms that integrate
the overdamped equation (\ref{eq:overdamped_Langevin}) by instead
directly discretizing the inertial equation (\ref{eq:inertial_Langevin})
without the inertial term $\partial_{t}\V v$. By carefully constructing
the corrector stage we will obtain the correct stochastic or thermal
drift term in (\ref{eq:overdamped_Langevin}). For the simple Langevin
equation (\ref{eq:inertial_Langevin}) a trapezoidal integrator goes
from time step $n$ to time step $n+1$ via the two stages,
\begin{align}
\M{\gamma}^{n}\V v^{n}= & \V F^{n}+\sqrt{\frac{2k_{B}T}{\D t}}\ \left(\M{\gamma}^{n}\right)^{\frac{1}{2}}\V W^{n}\nonumber \\
\V x^{p,n+1}= & \V x^{n}+\V v^{n}\D t\quad\mbox{(predictor)}\label{eq:pred_Lang}\\
\M{\gamma}^{p,n+1}\V v^{p,n+1}= & \V F^{p,n+1}+\sqrt{\frac{2k_{B}T}{\D t}}\ \left(\M{\gamma}^{n}\right)^{\frac{1}{2}}\V W^{n}\nonumber \\
\V x^{n+1}= & \V x^{n}+\left(\frac{\V v^{n}+\V v^{p,n+1}}{2}\right)\D t\quad\mbox{(corrector)},\label{eq:Fixman_Lang}
\end{align}
where superscripts denote the point at which a given quantity is evaluated,
for example, $\M{\gamma}^{p,n+1}=\M{\gamma}\left(\V x^{p,n+1}\right)$,
and $\V W^{n}$ is a collection of independent and identically distributed
(i.i.d.) standard normal random variables sampled independently at
each time step. This predictor-corrector method is only first-order
accurate for the multiplicative-noise case. When one is interested
in linearized equations (e.g., a particle trapped by a harmonic potential
to remain close to a stable minimum of the potential) the predictor-corrector
schemes we construct in section \ref{sec:Overdamped} are second-order
weakly accurate. It is not hard to show that the scheme (\ref{eq:Fixman_Lang})
is equivalent to the well-known Fixman integrator for (\ref{eq:overdamped_Langevin})
\cite{BD_Fixman}.

Note that the final Fixman update can be written in the form
\begin{eqnarray}
\V x^{n+1} & = & \V x^{n}+\left(\frac{\M M^{n}\V F^{n}+\M M^{p,n+1}\V F^{p,n+1}}{2}\right)\D t+\sqrt{2k_{B}T\D t}\,\left(\M M^{n}\right)^{\frac{1}{2}}\V W^{n},\nonumber \\
 & + & \D t\left(k_{B}T\right)\left(\M M^{p,n+1}-\M M^{n}\right)\left(2k_{B}T\D t\,\M M^{n}\right)^{-\frac{1}{2}}\V W^{n},\label{eq:RFD_Fixman}
\end{eqnarray}
where the mobility $\M M=\M{\gamma}^{-1}$. The first line is seen
as an application of the (second-order) explicit trapezoidal rule
for the deterministic drift terms and the Euler-Maryama method for
the stochastic terms, while the second line gives the thermal drift
term $\D t\left(k_{B}T\right)\partial_{\V x}\cdot\Mob(\V x)$ in expectation,
as we explain next.

\subsubsection{\label{sub:RFD}Random Finite Difference}

A key feature of the Fixman method is that it requires access to the
action of both $\M{\gamma}^{-1}$ and $\M{\gamma}^{\frac{1}{2}}$,
or equivalently, $\M{\gamma}$ and $\M{\gamma}^{-\frac{1}{2}}$. This
is a great disadvantage in cases when only the action of the mobility
$\M M=\M{\gamma}^{-1}$ and its factor $\M M^{\frac{1}{2}}$ are easily
computable \cite{BrownianBlobs}. An alternative method to obtain
the drift term $\left(k_{B}T\right)\partial_{\V x}\cdot\M M\left(\V x\right)$
in expectation is to use the general relation, 
\begin{equation}
\lim_{\delta\rightarrow0}\frac{1}{\delta}\av{\left(\Mob\left(\V x+\delta\D{\V x}\right)-\Mob\left(\V x\right)\right)\D{\V p}}=\partial_{\V x}\cdot\Mob(\V x),\label{eq:DriftWithIncrements}
\end{equation}
where $\D{\V x}$ and $\D{\V p}$ are Gaussian variates with mean
zero and covariance $\av{\D{\V x}_{i}\D{\V p}_{j}}=\delta_{ij}$.
The second line in the Fixman method (\ref{eq:RFD_Fixman}) can be
seen as an application of (\ref{eq:DriftWithIncrements}) with $\delta=\sqrt{\D t}$
and $\D{\V x}=\left(2k_{B}T\,\Mob^{n}\right)^{\frac{1}{2}}\M W^{n}$
and $\D{\V p}=\left(2k_{B}T\,\Mob^{n}\right)^{-\frac{1}{2}}\M W^{n}$.
The choice $\D{\V x}=\D{\V p}$ is, however, much simpler to use than
the Fixman choice because it does not require the application of either
$\left(\Mob^{n}\right)^{\frac{1}{2}}$ or $\left(\Mob^{n}\right)^{-\frac{1}{2}}$.
Here $\delta$ is a small discretization parameter that can be taken
to be related to $\D t$ as in the Fixman method, but this is not
necessary. One can more appropriately think of (\ref{eq:DriftWithIncrements})
as a ``random finite difference'' (RFD) with $\delta$ representing
the small spacing for the finite difference, to be taken as small
as possible while avoiding numerical roundoff problems. The advantage
of the ``random'' over a traditional finite difference is that only
a small number of evaluations of the mobility per time step is required.
Note that the subtraction of $\Mob\left(\V x\right)\D{\V p}$ in (\ref{eq:DriftWithIncrements})
is necessary in order to control the variance of the RFD estimate.

For the simple Langevin equation (\ref{eq:inertial_Langevin}) an
RFD approach uses the same predictor (\ref{eq:pred_Lang}) as in the
Fixman algorithm, but now with corrector, 
\begin{align}
\V v^{p,n+1}= & \left(\M{\gamma}^{p,n+1}\right)^{-1}\V F^{p,n+1}+\sqrt{\frac{2k_{B}T}{\D t}}\ \left(\M{\gamma}^{n}\right)^{-\frac{1}{2}}\V W^{n}\nonumber \\
\V x^{n+1}= & \V x^{n}+\left(\frac{\V v^{n}+\V v^{p,n+1}}{2}\right)\D t\nonumber \\
+ & \left(k_{B}T\right)\frac{\D t}{\delta}\left[\M{\gamma}^{-1}\left(\V x^{n}+\delta\,\widetilde{\M W}^{n}\right)-\M{\gamma}^{-1}\left(\V x^{n}\right)\right]\widetilde{\M W}^{n},\label{eq:RFD_Lang}
\end{align}
where $\widetilde{\M W}^{n}$ is a collection of independent standard
normal variates that are uncorrelated with $\V W^{n}$. Because the
corrector stage for the velocity evaluates the noise amplitude at
the beginning of the time step there are no stochastic drift terms
generated from the term $\left(\V v^{n}+\V v^{p,n+1}\right)\D t/2$,
and the RFD in the last line of (\ref{eq:RFD_Lang}) is necessary
to generate the missing drift in expectation.

It is not hard to see that (\ref{eq:RFD_Lang}) is a Markov chain
that is consistent with the Ito overdamped equation (\ref{eq:overdamped_Langevin}).
To demonstrate consistency we only need to show that to leading order,
the first moment of the increment is equal to the deterministic drift
in expectation, 
\[
\lim_{\D t\rightarrow0}\frac{1}{\D t}\, E\left(\V x^{n+1}-\V x^{n}\right)=\M M^{n}\V F^{n}+\left(k_{B}T\right)\partial_{\V x}\cdot\Mob(\V x^{n}),
\]
which follows directly from (\ref{eq:DriftWithIncrements}), and that
in expectation the second moment of the increment matches the covariance
of the noise,
\[
\lim_{\D t\rightarrow0}\frac{1}{\D t}\, E\left(\left(\V x^{n+1}-\V x^{n}\right)\left(\V x^{n+1}-\V x^{n}\right)^{T}\right)=2k_{B}T\,\M M^{n},
\]
which is trivially true because the random increment in both the predictor
and corrector stages is $\sqrt{2k_{B}T\D t}\,\left(\M M^{n}\right)^{\frac{1}{2}}\V W^{n}$.

The two schemes (\ref{eq:pred_Lang},\ref{eq:Fixman_Lang}) and (\ref{eq:pred_Lang},\ref{eq:RFD_Lang})
do not, of course, exhaust all possibilities. For example, in the
corrector stage for velocity we could evaluate the noise amplitude
at the predicted value,
\[
\V v^{p,n+1}=\left(\M{\gamma}^{p,n+1}\right)^{-1}\V F^{p,n+1}+\sqrt{\frac{2k_{B}T}{\D t}}\ \left(\M{\gamma}^{p,n+1}\right)^{-\frac{1}{2}}\V W^{n}.
\]
It is not difficult to show, however, that with this choice the term
$\left(\V v^{n}+\V v^{p,n+1}\right)\D t/2$ in the corrector stage
for the position would generate an incorrect stochastic drift term
for non-scalar problems. Therefore, additional RFD terms would be
required to remove any spurious drift terms and add the correct ones.
In this work we construct several schemes for integrating two-scale
systems such as (\ref{eq:overdamped_Langevin}) that generate the
correct drift terms using random finite differences, and, furthermore,
also obtain second-order weak accuracy for the overdamped equations
linearized around a stable deterministic trajectory. With additional
effort, it is also possible to obtain second-order weak accuracy for
the nonlinear overdamped equation by using weak Runge-Kutta schemes
of the kind developed in \cite{WeakSecondOrder_RK}, which also rely
on an RFD-like approach to avoid explicit evaluation of derivatives.

\subsection{\label{sub:FHD}Fluctuating Hydrodynamics}

As a motivating example of a fluctuating hydrodynamic system of equations
to which our methods will be applied, we study a model of diffusion
of tagged (labeled) molecules in a liquid or of colloidal particles
suspended in a fluid. A detailed mesoscopic model for diffusion in
liquids has been described by some of us in previous work \cite{DiffusionJSTAT,DDFT_Hydro};
here we only summarize some key points to give a specific setting
for the discussion to follow. The hydrodynamic fluctuations of the
fluid velocity $\V v\left(\V r,t\right)$ will be modeled via the
incompressible fluctuating Navier-Stokes equation, $\grad\cdot\V v=0$,
and
\begin{equation}
\rho\left(\partial_{t}\V v+\V v\cdot\grad\V v\right)+\grad\pi=\eta\grad^{2}\V v+\grad\cdot\left(\sqrt{2\eta k_{B}T}\,\M{\mathcal{W}}\right)-\beta\rho c\,\V g,\label{eq:fluct_NS}
\end{equation}
and appropriate boundary conditions. Here $\rho$ is the fluid density,
$\eta=\rho\nu$ the shear viscosity, and $T$ the temperature, all
assumed to be constant throughout the domain, and $\pi\left(\V r,t\right)$
is the mechanical pressure. The last term in this equation models
the effects of gravity using a Boussinesq constant density approximation,
with $\V g$ being the gravitational acceleration, and $\beta$ the
solutal expansion coefficient. The concentration of diffusing particles
is given by the mass fraction $c\left(\V r,t\right)=\left(m/\rho\right)n\left(\V r,t\right)$,
where $n$ is the number density and $m$ is the mass of the tracer
particles. The stochastic momentum flux is modeled via a white-noise
symmetric tensor field $\M{\mathcal{W}}\left(\V r,t\right)$ with
covariance chosen to obey a fluctuation-dissipation principle \cite{FluctHydroNonEq_Book,OttingerBook},
\[
\av{\mathcal{W}_{ij}(\V r,t)\mathcal{W}_{kl}(\V r^{\prime},t^{\prime})}=\left(\delta_{ik}\delta_{jl}+\delta_{il}\delta_{jk}\right)\delta(t-t^{\prime})\delta(\V r-\V r^{\prime}).
\]
Note that because the noise is additive in (\ref{eq:fluct_NS}) there
is no difference between an Ito and a Stratonovich interpretation
of the stochastic term. For technical reasons, in this work we will
drop the nonlinear advective term $\V v\cdot\grad\V v$ and consider
the time-dependent fluctuating Stokes equation; this is a good approximation
since this term only plays an important role at very large scales.

For our purposes, we will model the evolution of the concentration
$c\left(\V r,t\right)$ via a fluctuating advection-diffusion Ito
equation \cite{SPDE_Diffusion_Dean,DDFT_Hydro},
\begin{equation}
\partial_{t}c+\V u\cdot\grad c=\chi_{0}\grad^{2}c+\grad\cdot\left(\sqrt{2\chi_{0}\rho^{-1}mc}\,\M{\mathcal{W}}_{c}\right),\label{eq:c_eq_original}
\end{equation}
where $\M{\mathcal{W}}_{c}\left(\V r,t\right)$ denotes a white-noise
vector field. Here $\chi_{0}$ is a \emph{bare} or \emph{molecular}
diffusion coefficient, and the concentration is advected by the random
field 
\begin{equation}
\V u\left(\V r,t\right)=\int\M{\sigma}\left(\V r,\V r^{\prime}\right)\V v\left(\V r^{\prime},t\right)d\V r^{\prime}\equiv\M{\sigma}\star\V v,\label{eq:smoothing_u}
\end{equation}
where $\M{\sigma}$ is a smoothing kernel that filters out features
at scales below a microscopic (molecular) scale $\sigma$, and $\star$
denotes convolution. It is important here that $\V u$ is also divergence-free,
$\grad\cdot\V u=0$. Formally, one often writes the advective term
in (\ref{eq:c_eq_original}) as $\V v\cdot\grad c$ (this corresponds
to $\M{\sigma}\left(\V r,\V r^{\prime}\right)=\delta\left(\V r-\V r^{\prime}\right)$)
but this only makes sense if one truncates the velocity equation (\ref{eq:fluct_NS})
at some ultraviolet cutoff wave number $k_{\text{max}}\sim\sigma^{-1}$;
this kind of implicit smoothing kernel is applied by the finite-volume
spatial discretization we employ \cite{DFDB}, with the grid spacing
playing the role of $\sigma$. Additional filtering may also be implemented
in the finite-volume schemes, as described in Appendix B in \cite{LowMachExplicit}.
The system of equations (\ref{eq:fluct_NS},\ref{eq:c_eq_original})
is a useful model, for example, for the study of giant concentration
fluctuations in low-density polymer \cite{FractalDiffusion_Microgravity}
or nanocolloidal suspensions \cite{GiantFluct_NanoColloids}, or in
binary fluid mixtures in the presence of a modest temperature gradient
\cite{SoretDiffusion_Croccolo}.

It can be shown that the coupled velocity-concentration system (\ref{eq:fluct_NS},\ref{eq:c_eq_original})
can formally be written as an infinite-dimensional system of the form
(\ref{eq:x_t_general}) with a physically-sensible coarse-grained
energy \emph{functional} 
\begin{equation}
U\left[\V v\left(\cdot\right),\, c\left(\cdot\right)\right]=\frac{\rho}{2}\int v^{2}\left(\V r\right)\, d\V r+\beta\rho\int c\left(\V r\right)\left(\V r\cdot\V g\right)\, d\V r+\left(k_{B}T\right)\int n\left(\V r\right)\left(\ln\left(\Lambda^{3}n\left(\V r\right)\right)-1\right)\, d\V r,\label{eq:U_FNS}
\end{equation}
where $n=\rho c/m$ is the number density of the diffusing particles,
and $\Lambda$ is a fixed length-scale (e.g., the thermal de Broglie
wavelength). A specific form of the mobilty \emph{operator} can also
be written, see for example Refs. \cite{DFDB} for the advective terms
and Ref. \cite{DDFT_Hydro} for the diffusive and stochastic terms.
Numerical methods for integrating systems such (\ref{eq:fluct_NS},\ref{eq:c_eq_original})
as have been discussed in Refs. \cite{LLNS_Staggered,LowMachExplicit,DFDB}.
In particular, after spatial discretization of the system of SPDEs
(\ref{eq:fluct_NS},\ref{eq:c_eq_original}) one obtains a system
of SODEs of the generic Langevin form (\ref{eq:x_t_general}). We
note that the last term in (\ref{eq:U_FNS}), which contains the free
energy density of an ideal gas with number density $c\left(\V r\right)$,
is quite formal and poses notable mathematical (and numerical) difficulties
\cite{DDFT_Hydro}. In this work we will make a Gaussian approximation
and use instead the Gaussian free energy $k_{B}T/\left(2\vartheta_{0}\right)\int c^{2}\, d\V r$
\cite{DFDB}, where $\vartheta_{0}=\rho^{-1}mc_{0}$ and $c_{0}$
is the average concentration; this change simplifies the mobility
to a constant matrix and the noise term in (\ref{eq:c_eq_original})
becomes additive, $\grad\cdot\left(\sqrt{2\chi_{0}\vartheta_{0}}\,\M{\mathcal{W}}_{c}\right)$,
see Section \ref{sub:ConstCoeff}.

In practice, the physical properties of the fluid, notably, the viscosity
and the diffusion coefficient, depend on the concentration. This is
crucial, for example, to model experiments on the development of giant
concentration fluctuations during the diffusive mixing of water and
glycerol \cite{GiantFluctuations_Cannell}, since the viscosity and
diffusion coefficient (by virtue of the Stokes-Einstein relation)
depend very strongly on the concentration. Assuming that the density
changes only weakly with concentration we can use a Boussinesq approximation.
This approximation essentially amounts to assuming that the two fluid
components have very similar density so that the fluid density $\rho$
can be considered constant; for a generalization that accounts for
the fact that density depends on concentration see the low Mach number
formulation in \cite{LowMachExplicit}. The fluctuating hydrodynamic
equations in the case of variable transport coefficients can formally
be written as 
\begin{eqnarray}
\rho\left(\partial_{t}\V v+\V v\cdot\grad\V v\right)+\grad\pi & = & \grad\cdot\left(\eta\left(c\right)\bar{\grad}\V v+\sqrt{2\eta\left(c\right)k_{B}T}\,\M{\mathcal{W}}\right)-\beta\rho c\,\V g\nonumber \\
\partial_{t}c+\V v\cdot\grad c & = & \grad\cdot\left(\chi\left(c\right)\grad c+\sqrt{2k_{B}T\rho^{-1}\,\chi\left(c\right)\mu_{c}^{-1}\left(c\right)}\,\M{\mathcal{W}}_{c}\right),\label{eq:variable_coeff}
\end{eqnarray}
where in general the specified temperature $T\left(\V r,t\right)$
may depend on position and time. Here $\bar{\grad}=\grad+\grad^{T}$
and $\mu_{c}=\partial\mu/\partial c$ is the derivative of the chemical
potential of the binary mixture \cite{LowMachExplicit}. We have omitted
stochastic or thermal drift terms since these are poorly understood
(in fact, mathematically they are ill-defined) and not important in
the linearized dynamics. Note that in the linearized equation there
is no distinction between the bare and the effective diffusion coefficient,
so $\chi(c)$ denotes the macroscopic diffusion coefficient of the
ensemble-averaged concentration \cite{DiffusionJSTAT,DDFT_Hydro}.
Similarly, in the linearized equations the nonlinear advective terms
become a sum of linear terms, and therefore in (\ref{eq:variable_coeff})
we can write $\V v\cdot\grad c$ instead of $\V u\cdot\grad c$ and
still give a precise meaning to the linearized equations.

\subsubsection{\label{sub:OverdampedFHD}Overdamped Limit}

One of the key difficulties in directly integrating the system (\ref{eq:fluct_NS},\ref{eq:c_eq_original})
is the fact that in liquids momentum diffuses much more rapidly than
does mass, i.e., the dynamics of the velocity is much faster than
that of concentration. We used this fact in \cite{DiffusionJSTAT}
to eliminate the fast velocity adiabatically (see Appendix A in \cite{DiffusionJSTAT}
for technical details) and obtained a limiting or overdamped equation
for the concentration that takes the form of a Stratonovich SPDE,
\begin{equation}
\partial_{t}c=-\V w\odot\grad c+\beta\rho\eta^{-1}\left(\M G_{\sigma}\star c\V g\right)\cdot\grad c+\chi_{0}\grad^{2}c+\grad\cdot\left(\sqrt{2\chi_{0}\rho^{-1}m\, c}\,\M{\mathcal{W}}_{c}\right),\label{eq:limiting_Strato}
\end{equation}
where $\odot$ denotes a Stratonovich dot product, and the advection
velocity $\V w\left(\V r,t\right)$ is white in time, with covariance
proportional to a Green-Kubo integral of the velocity auto-correlation
function, 
\begin{align}
\av{\V w\left(\V r,t\right)\otimes\V w\left(\V r^{\prime},t^{\prime}\right)} & =2\delta\left(t-t^{\prime}\right)\int_{0}^{\infty}\av{\V u\left(\V r,t\right)\otimes\V u\left(\V r^{\prime},t+t^{\prime\prime}\right)}dt^{\prime\prime}=\label{eq:C_w}\\
 & =\frac{2k_{B}T}{\eta}\delta\left(t-t^{\prime}\right)\int\M{\sigma}\left(\V r,\V r^{\prime\prime}\right)\M G\left(\V r^{\prime\prime},\V r^{\prime\prime\prime}\right)\M{\sigma}^{T}\left(\V r^{\prime},\V r^{\prime\prime\prime}\right)d\V r^{\prime\prime}d\V r^{\prime\prime\prime}.
\end{align}
Here $\M G$ denotes the Green's function for Stokes flow, and $\M G_{\sigma}$
denotes $\M G$ regularized by the smoothing kernel $\M{\sigma}$;
more explicitly, $\M G_{\sigma}\star\V f$ is a shorthand notation
for the smoothed solution of the Stokes equation with unit viscosity:
$\V w_{\sigma}=\M G_{\sigma}\star\V f$ if $\V w_{\sigma}=\M{\sigma}\star\V w$
and $\V w=\M G\star\V f$ solves 
\begin{equation}
\grad\pi=\grad^{2}\V w+\V f,\qquad\grad\cdot\V w=0,\label{eq:22}
\end{equation}
along with appropriate boundary conditions.

In this paper we will describe an algorithm that can be used to integrate
the overdamped limit (\ref{eq:limiting_Strato}) with a time step
size several orders of magnitude larger than the time step size required
to integrate the original inertial dynamics (\ref{eq:fluct_NS},\ref{eq:c_eq_original}).
The spatial discretization will be the same as for the original system
and only the temporal integrator will change. We first described such
a temporal integrator that \emph{automatically} performs adiabatic
mode elimination \emph{without} directly discretizing or even writing
the limiting dynamics in Appendix B of \cite{DiffusionJSTAT}. Here
we generalize this to a broad class of systems of Langevin SDEs that
contain a fast and a slow variable.

\subsubsection{\label{sub:LFHD}Linearized Fluctuating Hydrodynamics}

At microscopic scales, the nonlinearity and Stratonovich nature of
the advective term $\V w\odot\grad c$ in (\ref{eq:limiting_Strato})
is crucial. This advection by the rapidly fluctuating random velocity
field contributes to the effective diffusion, similar to eddy diffusivity
in turbulent flows. In particular, the ensemble averaged concentration
$\bar{c}\left(\V r,t\right)=\av{c\left(\V r,t\right)}$ is described
by Fick's macroscopic law with a renormalized diffusion coefficient
\begin{equation}
\partial_{t}\bar{c}=\grad\cdot\left[\left(\chi_{0}+\M{\chi}\right)\grad\bar{c}\right]=\grad\cdot\left(\M{\chi}_{\text{eff}}\grad\bar{c}\right),\label{eq:Fick_law}
\end{equation}
where the fluctuation-induced diffusion tensor $\M{\chi}\left(\V r\right)$
is given by a Green-Kubo formula and follows a Stokes-Einstein relation
\cite{DiffusionJSTAT}. This enhancement of the diffusion coefficient
is mathematically a stochastic drift term that comes from the divergence
of the mobility operator (last term in (\ref{eq:x_t_general})) when
one writes (\ref{eq:limiting_Strato}) in the Ito formulation \cite{DiffusionJSTAT}.
Our temporal integrators will capture this term using a predictor-corrector
algorithm, without explicitly evaluating derivatives.

At mesoscopic length-scales $\delta\gg\sigma$ much larger than the
molecular, in three dimensions, one expects that the fluctuations
in $c{}_{\delta}=\bar{c}+\d c$, where $\bar{c}=\av c$ is the solution
of the \emph{deterministic} Fick's law (\ref{eq:Fick_law}), are small
and approximately Gaussian (this is a form of a law of large numbers
and a central limit theorem for the fluctuations). In particular,
a widely-used model for the fluctuations in the concentration at such
scales is \emph{linearized fluctuating hydrodynamics }\cite{FluctHydroNonEq_Book}
(LFH). In LFH, one first solves the macroscopic \emph{deterministic}
hydrodynamic equations first, and then linearizes the formal nonlinear
fluctuating hydrodynamic equations to leading order in the fluctuations.
If we assume that there is no macroscopic convective motion of the
fluid, so that the solution of the deterministic Navier-Stokes equation
$\bar{\V v}=0$ and $\d{\V v}\equiv\V v$, the LFH equations become
\begin{align}
\partial_{t}\bar{c} & =\chi_{\text{eff}}\grad^{2}\bar{c}\nonumber \\
\partial_{t}\left(\d c\right) & =-\V v\cdot\grad\bar{c}+\chi_{\text{eff}}\grad^{2}\left(\d c\right)+\grad\cdot\left(\sqrt{2\chi_{\text{eff}}\rho^{-1}m\,\bar{c}}\,\V{\mathcal{W}}_{c}\right)\label{eq:linearized_diffusion}\\
\rho\partial_{t}\V v+\grad\pi & =\eta\grad^{2}\V v+\grad\cdot\left(\sqrt{2\eta k_{B}T}\,\M{\mathcal{W}}\right)-\beta\rho\left(\d c\right)\,\V g.\nonumber 
\end{align}
In the first part of this paper we describe how to numerically solve
these types of linearized Langevin equations \emph{without} solving
the deterministic equations first and then explicitly linearizing
around them. Specifically, we will construct temporal integrators
that perform the linearization \emph{numerically}.

Note that in cases when there is a large separation of time scales
between the velocity and the concentration one may perform adiabatic
elimination of the velocity; in the linearized setting this simply
amounts to dropping the inertial term $\rho\partial_{t}\V v$ and
switching to the time-independent Stokes equation for the velocity.
Formally, the linearized limiting or overdamped equation is 
\begin{eqnarray}
\partial_{t}\bar{c} & = & \chi_{\text{eff}}\grad^{2}\bar{c}\label{eq:linearized_overdamped}\\
\partial_{t}\left(\d c\right) & = & \beta\rho\eta^{-1}\left(\M G\star\left(\d c\right)\V g\right)\cdot\grad\bar{c}+\left(\M G_{\myhalf}\sqrt{2\eta^{-1}k_{B}T}\,\M{\mathcal{W}}\right)\cdot\grad\bar{c}\nonumber \\
 & + & \chi_{\text{eff}}\grad^{2}\left(\d c\right)+\grad\cdot\left(\sqrt{2\chi_{\text{eff}}\rho^{-1}m\,\bar{c}}\,\V{\mathcal{W}}_{c}\right),\nonumber 
\end{eqnarray}
where $\M G$ is the Green's function for steady Stokes flow and $\M G_{\myhalf}$
symbolically denotes that the covariance of the additive-noise term
$\M G_{\myhalf}\M{\mathcal{W}}$ is $\M G$.

In our previous work \cite{DFDB} we described several semi-implicit
predictor-corrector temporal integrators for the inertial system (\ref{eq:fluct_NS},\ref{eq:c_eq_original}),
which has \emph{time-independent} additive noise. In this paper we
will describe a simple modification that allows one to numerically
linearize and then integrate, with second-order weak accuracy, the
linearization (\ref{eq:linearized_diffusion}) around the \emph{time-dependent}
solution of the \emph{nonlinear} deterministic equations. We will
also construct a \emph{single} unified numerical method that can be
used to integrate either the nonlinear (\ref{eq:limiting_Strato})
(microscopic scales, large noise) or the linearized (\ref{eq:linearized_overdamped})
(mesoscopic scales, weak noise) overdamped equations. Which equation
is appropriate depends sensitively on the spatial scale of interest,
specifically, on the range of wavenumbers whose dynamics needs to
be captured accurately.

In this work we will develop temporal integrators suitable also for
the more general variable-coefficient equations (\ref{eq:variable_coeff}).
Since our temporal integrators will work with the original equations
but in the end simulate the correct overdamped or linearized dynamics,
we will never need to explicitly write down the (more complex) overdamped
limit or the linearized system. We simply let the numerical method
do that for us.

\section{\label{sec:Linearized}Temporal Integrators for Linearized Langevin
Equations}

In this section we consider a relatively general system of Langevin
equations for the coarse-grained variable $\V x\left(t\right)$,
\begin{equation}
\frac{d\V x}{dt}=\V f(\V x)+\M K(\V x)\diamond\V{\mathcal{W}}\left(t\right)=\M H(\V x)\V x+\V h(\V x)+\M K(\V x)\diamond\V{\mathcal{W}}\left(t\right),\label{eq:dx_dt_nonlin}
\end{equation}
where $\M H(\V x)\V x$ is a term that we may choose to treat semi-implicitly
in cases when it is stiff, and $\V h(\V x)$ denotes the remaining
terms which are difficult to treat implicitly. In cases when the noise
is multiplicative one must choose a specific interpretation of the
stochastic integral, here we have chosen the kinetic product $\diamond$
\cite{KineticStochasticIntegral_Ottinger} in agreement with our motivating
equation (\ref{eq:x_t_kinetic}). The system (\ref{eq:dx_dt_nonlin})
may arise from a spatial discretization of fluctuating hydrodynamics
SPDEs, but similar equations arise in a variety of contexts.

Our focus will be on developing methods for integrating the system
of equations obtained after linearizing (\ref{eq:dx_dt_nonlin}) around
the solution $\bar{\V x}(t)$ of the deterministic system of equations
(obtained by simply dropping the noise term),
\begin{align}
d\bar{\V x}/dt= & \M H(\bar{\V x})\bar{\V x}+\V h(\bar{\V x})\label{eq:dxbar_dt}\\
d(\delta\V x)/dt= & \M M(\bar{\V x})\delta\V x+\M K(\bar{\V x})\V{\mathcal{W}}\left(t\right),\label{eq:dx_dt_lin}
\end{align}
where $\delta\V x=\V x-\bar{\V x}$ is the (presumably small) fluctuation
around the deterministic dynamics. Here the Jacobian of $\V f(\V x)$
is denoted with 
\[
\M M(\bar{\V x})=\partial_{\V x}\V f(\bar{\V x})=\M H\left(\bar{\V x}\right)+\left(\partial_{\V x}\M H\left(\bar{\V x}\right)\right)\bar{\V x}+\partial_{\V x}\V h\left(\bar{\V x}\right),
\]
more specifically, in index notation
\[
M_{ij}=H_{ij}\left(\bar{\V x}\right)+\left(\partial_{j}H_{ik}\left(\bar{\V x}\right)\right)\bar{x}_{k}+\partial_{j}h_{i}\left(\bar{\V x}\right).
\]
Note that the noise in (\ref{eq:dx_dt_lin}) is time-dependent but
still additive, and different interpretations of the stochastic integral
are equivalent.

One can more precisely justify the system (\ref{eq:dx_dt_lin}) by
assuming that the noise is very weak. In particular, the deterministic
equation (\ref{eq:dxbar_dt}) can be seen as a law of large numbers
describing the most probable trajectory in the weak-noise limit, with
the Ornstein-Uhlenbeck equation (\ref{eq:dx_dt_lin}) as a central-limit
theorem for the small nearly Gaussian fluctuations around the average.
It is important to note that one must assume here that the deterministic
dynamics is stable, that is, small perturbations do not lead to large
deviations of the averages, which is a good assumption far from phase
transitions or bifurcation points. Note, however, that the linearized
equations cannot be used to describe rare events (large deviations)
or events that occur on exponentially-long timescales.

The essential difficulty in integrating (\ref{eq:dx_dt_lin}) directly
is that the linearization needs to be performed around a time-dependent
state $\bar{\V x}(t)$ that is not known \emph{a priori} but is rather
itself the solution of a nonlinear system of equations. Furthermore,
one must calculate the Jacobian $\M M$ explicitly, and this is often
quite tedious since many more terms appear in the linearization than
do in the original nonlinear equations (this is especially true for
fluctuating hydrodynamics). Instead, we will construct methods that
directly work with the original nonlinear equation (\ref{eq:dx_dt_nonlin})
but with a noise term that is deliberately made very weak, as some
of us first proposed and applied in \cite{LLNS_Staggered} to a case
of a \emph{steady} deterministic state. In fact, if the assumption
of weak noise used to justify the linearization is actually correct,
the noise does not have to be artificially reduced in magnitude at
all and using the actual (physical) value of the noise amplitude will
give indistinguishable results. While it is of course always better
to simply integrate the original nonlinear dynamics in cases where
it is known, it is important to emphasize that nonlinear fluctuating
hydrodynamics is very poorly understood and in most cases the nonlinear
equations are ill-posed; a notable exception are (\ref{eq:fluct_NS},\ref{eq:c_eq_original})
and (\ref{eq:limiting_Strato}) because the nonlinear advective term
there was carefully regularized in a physically-relevant manner \cite{DiffusionJSTAT}.
By contrast, the linearized equations are well-defined because there
are no nonlinear terms and a precise meaning can be given in the space
of (Gaussian) distributions \cite{DaPratoBook}. Another important
reason for focusing on the linearized equations is that while integrating
the nonlinear equations to second-order (weakly) is rather nontrivial
in the case of multiplicative noise \cite{WeakSecondOrder_RK}, it
is not hard to construct simple second-order integrators for the linearized
equations, as we demonstrate here.

Before we describe methods for solving (\ref{eq:dxbar_dt},\ref{eq:dx_dt_lin}),
we describe how to construct temporal integrators for just (\ref{eq:dx_dt_lin}),
assuming that $\bar{\V x}(t)$ is known and given to us, and therefore
the noise amplitude $\M K(\bar{\V x}\left(t\right))\equiv\M K(t)$
is only a function of time.

\subsection{Time-dependent noise}

At first, it will not be necessary to assume that the equation for
the fluctuations is linear, and we will therefore consider a more
general SDE with time dependent additive noise, 
\begin{align}
\frac{d\V x}{dt} & =\M L(\V x)\V x+\V g(\V x)+\M K(t)\V{\mathcal{W}}\left(t\right),\label{eq:TimeDepNoise}
\end{align}
which is a generalization of the constant additive noise equation
considered in \cite{DFDB}. Such an equation may arise, for example,
by considering a time-dependent temperature in the fluctuating Navier-Stokes
equation (\ref{eq:fluct_NS}). Here $\V{\mathcal{W}}\left(t\right)$
denotes a collection of independent white-noise processes, formally
identified with the time derivative of a collection of independent
Brownian motions (Wiener processes) $\V{\mathcal{B}}\left(t\right)$,
$\V{\mathcal{W}}\equiv d\V{\mathcal{B}}/ds$, $\V g\left(\V x\right)$
denotes all of the terms handled explicitly (e.g., advection or external
forcing), and the term $\M L(\V x)\V x$ denotes terms that will be
handled semi-implicitly (e.g., diffusion) for stiff systems (large
spread in the eigenvalues of $\M L$). In general, $\V L\left(\V x\right)$
may depend on $\V x$ since the transport coefficients (e.g., viscosity)
may depend on certain state variables (e.g., concentration). Note
that the equation (\ref{eq:TimeDepNoise}) also includes the case
where $\M L(\V x,t)$ and $\V g(\V x,t)$ depend explicitly on time,
as can be seen by considering an expanded system of equations for
$\V x\to(\V x,t).$ 

Here we focus on weak integrators that are (at most) second-order
accurate. The following is a relatively general mixed explicit-implicit
predictor-corrector scheme for solving (\ref{eq:TimeDepNoise}) that
is a slight generalization of the scheme discussed in \cite{DFDB}.
The first stage in our schemes is a predictor step to estimate $\tilde{\V x}\approx\V x\left(n\D t+w_{2}\D t\right)$,
where $w_{2}$ is some chosen weight (e.g., $w_{2}=1/2$ for a midpoint
predictor or $w_{2}=1$ for a full-step predictor), while the corrector
stage completes the step by estimating $\V x^{n+1}$ at time $(n+1)\D t$,
\begin{align}
\V x^{(p)}= & \V x^{n}+\left((w_{2}-w_{1})\M L^{n}\V x^{n}+w_{1}\M L^{n}\V x^{(p)}\right)\D t+w_{2}\V g^{n}\D t+\sqrt{w_{2}\D t}\,\M K^{n}\V W^{n,1}\nonumber \\
\V x^{n+1}= & \V x^{n}+\left((1-w_{3}-w_{4}-w_{5})\M L^{n}\V x^{n}+w_{3}\M L^{(p)}\V x^{(p)}+w_{4}\M L^{(p)}\V x^{n+1}+w_{5}\M L^{n}\V x^{n+1}\right)\D t\nonumber \\
 & +\D t\;\begin{cases}
\left((1-w_{6})\V g^{n}+w_{6}\V g^{(p)}\right), & \mbox{or}\\
\V g\left((1-w_{6})\V x^{n}+w_{6}\V x^{(p)}\right)
\end{cases}\nonumber \\
 & +\left((1-w_{7})\M K^{n}+w_{7}\M K^{(p)}\right)\left(\sqrt{w_{2}\D t}\,\V W^{n,1}+\sqrt{(1-w_{2})\D t}\,\M W^{n,2}\right),\label{eq:predcorr_scheme}
\end{align}
where we have denoted $\M K^{(p)}=\M K(t_{n}+w_{2}\D t)$ and superscripts
and decorations denote the point at which a given quantity is evaluated,
for example, $\M L^{(p)}=\M L\left(\V x^{(p)}\right)$. Note that
this class of semi-implicit schemes requires solving only \emph{linear}
systems involving the matrix $\M L(\V x)$ evaluated at a specific
point and kept fixed. In the above discretization, the standard normal
variates $\V W_{1}^{n}$ correspond to the increment of the underlying
Wiener processes $\V{\mathcal{B}}(t)$ over the time interval $w_{2}\D t$,
$\V{\mathcal{B}}\left(n\D t+w_{2}\D t\right)-\V{\mathcal{B}}\left(n\D t\right)=\left(w_{2}\Delta t\right)^{\frac{1}{2}}\V W_{1}^{n}$
in law, while the normal variates $\V W_{2}^{n}$ correspond to the
independent increment over the remainder of the time step, $\V{\mathcal{B}}\left(\left(n+1\right)\D t\right)-\V{\mathcal{B}}\left(n\D t+w_{2}\D t\right)=\left(\left(1-w_{2}\right)\Delta t\right)^{\frac{1}{2}}\V W_{2}^{n}$
in law. We give two alternative ways to handle the explicit terms
in the corrector stage, which give the same order of accuracy, and,
are, in fact, identical if $\V g$ is linear. Which of the two ways
of handling the explicit terms is better should be tested empirically
for strongly nonlinear Langevin equations, as we did in Ref. \cite{DFDB}
using the stochastic Burgers equation as a model problem.

Different specific values for the $w$ coefficients determine different
schemes. The analysis summarized in Appendix A in \cite{DFDB} is
extended in Appendix \ref{sec:Appendix-time-dependent} to account
for the time dependence of $\M K$, and shows that the scheme (\ref{eq:predcorr_scheme})
is weakly second order accurate if the weights satisfy
\begin{align}
w_{3}w_{2}+w_{4}w_{2}= & \frac{1}{2}\nonumber \\
w_{3}w_{2}+w_{4}+w_{5}= & \frac{1}{2}\nonumber \\
w_{2}w_{6}= & \frac{1}{2}\label{eq:second_order_cond}\\
w_{2}w_{7}= & \frac{1}{2}.\nonumber 
\end{align}
We present two simple schemes that satisfy these properties next,
the first fully explicit, and the second semi-implicit. While these
by no means exhaust all possibilities, they are representative and
have several notable advantages for the case of time-independent noise,
as discussed in more detail in \cite{DFDB}.

\subsubsection{Explicit Midpoint Scheme}

A fully explicit midpoint predictor-corrector scheme is obtained for
$w_{1}=0,\, w_{2}=1/2,\, w_{3}=1,\, w_{4}=w_{5}=0,\, w_{6}=1,w_{7}=1$:
\begin{align}
\V x^{p,n+\frac{1}{2}}= & \V x^{n}+\frac{\D t}{2}\left(\M L^{n}\V x^{n}+\V g^{n}\right)+\sqrt{\frac{\D t}{2}}\M K^{n}\V W_{1}^{n}\nonumber \\
\V x^{n+1}= & \V x^{n}+\D t\left(\M L^{p,n+\frac{1}{2}}\V x^{p,n+\frac{1}{2}}+\V g^{p,n+\frac{1}{2}}\right)+\sqrt{\frac{\D t}{2}}\V K^{p,n+\frac{1}{2}}\left(\V W_{1}^{n}+\V W_{2}^{n}\right).\label{eq:expl_midpoint}
\end{align}
This midpoint scheme has several notable strengths for fluctuating
hydrodynamics:
\begin{enumerate}
\item It is fully explicit and thus quite efficient (but also subject to
restrictive stability limits on the time step size for stiff systems).
\item It is a weakly second-order accurate integrator for (\ref{eq:TimeDepNoise}).
\item It is third-order accurate for static structure factors (static correlations)
for time-independent additive noise in the linearized setting \cite{DFDB},
that is, for the equation
\begin{equation}
\frac{d\V x}{dt}=\M L\V x+\M K\,\V{\mathcal{W}}\left(t\right),\label{eq:const_noise}
\end{equation}
with constant $\M L$ and $\M K$.
\end{enumerate}
It is also possible to construct an explicit trapezoidal scheme that
only uses a single random increment per time step, $w_{1}=0,\, w_{2}=1,\, w_{3}=1/2,\, w_{4}=w_{5}=0,w_{6}=1/2,\, w_{7}=1/2$,
but for fluctuating hydrodynamics the explicit midpoint integrator
is preferred because it gives third-order accurate static correlations.

\subsubsection{Implicit Trapezoidal Integrator}

We obtain a semi-implicit trapezoidal predictor-corrector scheme for
$w_{1}=1/2,\, w_{2}=1,\, w_{3}=w_{5}=0,\, w_{4}=1/2,\, w_{6}=1/2,\, w_{7}=1/2$:
\begin{eqnarray}
\V x^{p,n+1} & = & \V x^{n}+\frac{\Delta t}{2}\M L^{n}\left(\V x^{n}+\V x^{p,n+1}\right)+\Delta t\,\V g^{n}+\sqrt{\D t}\,\M K^{n}\V W^{n}\nonumber \\
\V x^{n+1} & = & \V x^{n}+\frac{\Delta t}{2}\left(\M L^{n}\V x^{n}+\M L^{p,n+1}\V x^{n+1}\right)+\frac{\Delta t}{2}(\V g^{n}+\V g^{p,n+1})+\frac{\sqrt{\D t}}{2}\left(\M K^{n}+\M K^{p,n+1}\right)\V W^{n}.\label{eq:impl_trapezoidal}
\end{eqnarray}
This scheme has the following advantages:
\begin{enumerate}
\item It only requires solving two linear systems with the coefficient matrix
$\M I-\left(\D t/2\right)\M L$, which is quite standard in computational
fluid dynamics and can be done efficiently using multigrid techniques.
\item It is a weakly second-order accurate integrator for (\ref{eq:TimeDepNoise}).
\item It is stable and gives the \emph{exact} \emph{static} correlations
for (\ref{eq:const_noise}) for \emph{any} time step size $\D t$
\cite{DFDB}.
\end{enumerate}
The alternative handling of the explicit term leads to the corrector
stage,
\begin{equation}
\V x^{n+1}=\V x^{n}+\frac{\Delta t}{2}\left(\M L^{n}\V x^{n}+\M L^{p,n+1}\V x^{n+1}\right)+\D t\,\V g\left(\frac{\V x^{n}+\V x^{p,n+1}}{2}\right)+\frac{\sqrt{\D t}}{2}\left(\M K^{n}+\M K^{p,n+1}\right)\V W^{n},\label{eq:impl_trap_expl_mid}
\end{equation}
which has the same advantages as (\ref{eq:impl_trapezoidal}), but
may behave differently for strongly nonlinear equations.

\subsubsection{Multiplicative noise}

The scheme (\ref{eq:predcorr_scheme}) with the conditions (\ref{eq:second_order_cond})
is a second-order weak integrator for the additive-noise equation
(\ref{eq:TimeDepNoise}). With a simple addition of a random finite
difference (RFD) term in the corrector, see Section \ref{sub:RFD},
the scheme (\ref{eq:predcorr_scheme}) can be turned into a first-order
weak integrator for the nonlinear kinetic SDE 
\begin{equation}
\frac{d\V x}{dt}=\M L(\V x)\V x+\V g(\V x)+\M K(\V x,t)\diamond\V{\mathcal{W}}\left(t\right).\label{eq:mult_noise}
\end{equation}
Namely, if we interpret $\M K^{(p)}=\M K(\V x^{(p)},\, t_{n}+w_{2}\D t)$
as the noise amplitude evaluated at the predictor, the integrator
(\ref{eq:predcorr_scheme}) is consistent with a Stratonovich interpretation
of the noise. If we want the integrator to be consistent with the
kinetic interpretation, we need to add a missing piece of the stochastic
drift term,
\[
\frac{1}{2}\partial_{\V x}\cdot\left(\M K\M K^{\star}\right)=\frac{1}{2}\left(\partial_{\V x}\M K\right):\M K^{\star}+\frac{1}{2}\M K\left(\partial_{\V x}\cdot\M K^{\star}\right).
\]
The first term on the right hand side in index notation reads $\left(\partial_{j}K_{ik}\right)K_{jk}$
and is the only drift term that appears if a Stratonovich interpretation
of the noise is adopted. To obtain the second term, we need to add
to the corrected $\V x^{n+1}$ the following RFD increment,
\begin{equation}
\D{\V x^{n+1}}=\frac{\D t}{2\delta}\M K\left(\M K^{\star}\left(\V x^{n}+\delta\widetilde{\V W}^{n}\right)-\M K^{\star}\left(\V x^{n}\right)\right)\widetilde{\V W}^{n},\label{eq:RFD_kinetic}
\end{equation}
where $\widetilde{\V W}^{n}$ is a collection of i.i.d. standard normal
increments generated independently of $\V W^{n}$.

\subsection{\label{sub:Linearization}Linearization around complex deterministic
flows}

We now turn our attention to temporal integrators for the linearized
system (\ref{eq:dxbar_dt},\ref{eq:dx_dt_lin}). The key difference
with (\ref{eq:TimeDepNoise}) is that we do not assume that $\bar{\V x}(t)$
is known, rather, it is also obtained by the numerical method. Our
approach will be to pretend we are integrating the nonlinear equation
(\ref{eq:dx_dt_nonlin}) but using very weak noise, so that we will
effectively be integrating the linearized equation. Our goal will
be to construct schemes that are second-order weakly accurate for
the linearized system (\ref{eq:dxbar_dt},\ref{eq:dx_dt_lin}). For
steady states, that is, when the deterministic or background state
$\bar{\V x}$ is independent of time, the linearized equation (\ref{eq:dx_dt_lin})
is of the form (\ref{eq:TimeDepNoise}) with the identification $\V x\equiv\delta\V x$
and $\M K\left(t\right)\equiv\M K\left(\bar{\V x},t\right)$, along
with 
\begin{equation}
\M L\equiv\M H\left(\bar{\V x}\right)=\text{const.}\quad\mbox{and}\quad\V g\left(\V x\right)\equiv\left[\left(\partial_{\V x}\M H\left(\bar{\V x}\right)\right)\bar{\V x}+\partial_{\V x}\V h\left(\bar{\V x}\right)\right]\V x\label{eq:steady_L_g}
\end{equation}
as the part of the linear operator treated implicitly and explicitly,
respectively. We would like to construct one fully explicit integrator
that for steady states becomes equivalent to the explicit midpoint
scheme (\ref{eq:expl_midpoint}), and one semi-implicit integrator
that for steady states becomes equivalent to the implicit trapezoidal
scheme (\ref{eq:impl_trapezoidal}). In this way, we obtain simple
schemes that inherit all of the important strengths of the explicit
trapezoidal and implicit midpoint schemes in describing fluctuations
around a steady state, while also performing numerical linearization
and maintaining second-order weak accuracy for time-dependent problems.

\subsubsection{Explicit schemes}

Let us first consider fully explicit schemes, for which the analysis
is considerably easier. A simple predictor-corrector update for the
nonlinear equation (\ref{eq:dx_dt_nonlin}) reads 
\begin{align}
\V x^{(p)}= & \V x^{n}+w_{2}\D t\,\V f^{n}+\sqrt{w_{2}\D t}\,\M K^{n}\V W_{1}^{n}\quad\mbox{(predictor)}\label{eq:explicit_linearizer}\\
\V x^{n+1}= & \V x^{n}+\left(\left(1-w_{3}\right)\V f^{n}+w_{3}\V f^{(p)}\right)\D t\nonumber \\
 & +\left((1-w_{7})\M K^{n}+w_{7}\M K^{(p)}\right)\left(\sqrt{w_{2}\D t}\,\V W_{1}^{n}+\sqrt{(1-w_{2})\D t}\,\V W_{2}^{n}\right)\quad\mbox{(corrector)}.\nonumber 
\end{align}
If we now split the variables into deterministic and fluctuating components,
\begin{eqnarray}
\V x^{(p)} & = & \bar{\V x}^{(p)}+\delta\V x^{(p)}\nonumber \\
\V x^{n+1} & = & \bar{\V x}^{n+1}+\delta\V x^{n+1}\label{eq:mean_plus_fluct}
\end{eqnarray}
and expand the predictor and corrector stages to first order in the
fluctuations (i.e., linearize both stages in $\delta\V x^{(p)}$ and
$\delta\V x^{n+1}$), we obtain that for weak fluctuations (\ref{eq:explicit_linearizer})
is effectively applying the same explicit scheme to the linearized
system (\ref{eq:dxbar_dt},\ref{eq:dx_dt_lin}). The analysis summarized
in Appendix \ref{Appendix-Accuracy} shows that the scheme (\ref{eq:explicit_linearizer})
is a second-order weakly accurate integrator for (\ref{eq:dxbar_dt},\ref{eq:dx_dt_lin})
if $w_{2}w_{3}=1/2$ and $w_{2}w_{7}=1/2$. Examples of second-order
fully explicit schemes include the explicit midpoint scheme, obtained
for $w_{2}=1/2,\, w_{3}=1,\, w_{7}=1$, and the explicit trapezoidal
scheme, obtained for $w_{2}=1,\, w_{3}=1/2,\, w_{7}=1/2$. Note that
at steady state the explicit midpoint scheme becomes equivalent to
applying (\ref{eq:expl_midpoint}) directly to (\ref{eq:dx_dt_lin}).

\subsubsection{Semi-implicit schemes}

We now consider applying the relatively general predictor-corrector
scheme (\ref{eq:predcorr_scheme}) to (\ref{eq:dx_dt_nonlin}), with
the identification $\M L\left(\V x\right)\equiv\M H\left(\V x\right)$
and $\V g(\V x)\equiv\V h(\V x)$, which gives the semi-implicit scheme
\begin{align}
\V x^{(p)}= & \V x^{n}+\left((w_{2}-w_{1})\M H^{n}\V x^{n}+w_{1}\M H^{n}\V x^{(p)}\right)\D t+w_{2}\V h^{n}\D t+\sqrt{w_{2}\D t}\,\M K^{n}\V W_{1}^{n}\nonumber \\
\V x^{n+1}= & \V x^{n}+\left((1-w_{3}-w_{4}-w_{5})\M H^{n}\V x^{n}+w_{3}\M H^{(p)}\V x^{(p)}+w_{4}\M H^{(p)}\V x^{n+1}+w_{5}\M H^{n}\V x^{n+1}\right)\D t\nonumber \\
 & +\begin{cases}
\left((1-w_{6})\V h^{n}+w_{6}\V h^{(p)}\right)\D t, & \mbox{or}\\
\V h\left((1-w_{6})\V x^{n}+w_{6}\V x^{(p)}\right)\D t
\end{cases}\label{eq:general_linearizer}\\
 & +\left((1-w_{7})\M K^{n}+w_{7}\M K^{(p)}\right)\left(\sqrt{w_{2}\D t}\,\V W_{1}^{n}+\sqrt{(1-w_{2})\D t}\,\V W_{2}^{n}\right).\nonumber 
\end{align}
If we substitute (\ref{eq:mean_plus_fluct}) in (\ref{eq:general_linearizer})
and linearize both stages, we obtain that for weak noise the same
scheme (but without the noise terms) is applied to the deterministic
component,
\begin{align}
\V{\bar{x}}^{(p)}= & \V{\bar{x}}^{n}+\left((w_{2}-w_{1})\M{\bar{H}}^{n}\V{\bar{x}}^{n}+w_{1}\M{\bar{H}}^{n}\V{\bar{x}}^{(p)}\right)\D t+w_{2}\V{\bar{h}}^{n}\D t\nonumber \\
\V{\bar{x}}^{n+1}= & \V{\bar{x}}^{n}+\left((1-w_{3}-w_{4}-w_{5})\M{\bar{H}}^{n}\V{\bar{x}}^{n}+w_{3}\M{\bar{H}}^{(p)}\V{\bar{x}}^{(p)}+w_{4}\M{\bar{H}}^{(p)}\V{\bar{x}}^{n+1}+w_{5}\M{\bar{H}}^{n}\V{\bar{x}}^{n+1}\right)\D t\nonumber \\
 & +\left((1-w_{6})\V{\bar{h}}^{n}+w_{6}\V{\bar{h}}^{(p)}\right)\D t\mbox{ or }\V h\left((1-w_{6})\V{\bar{x}}^{n}+w_{6}\V{\bar{x}}^{(p)}\right)\D t.\label{eq:general_x_bar}
\end{align}
while for the fluctuating component, in index notation, 
\begin{align}
\delta x_{i}^{(p)}= & \delta x_{i}^{n}+(w_{2}-w_{1})\left(\bar{H}_{ij}^{n}+\left(\partial_{j}\bar{H}_{ik}^{n}\right)\bar{x}_{k}^{n}\right)\delta x_{j}^{n}\D t\label{eq:general_dx_p}\\
+ & \left(w_{1}\left(\partial_{k}\bar{H}_{ij}^{n}\right)\bar{x}_{j}^{(p)}\delta x_{k}^{(n)}+w_{1}\bar{H}_{ij}\delta x_{j}^{(p)}\right)\D t\nonumber \\
+ & w_{2}\left(\partial_{j}\bar{h}_{i}\right)\delta x_{j}^{n}\D t+\sqrt{w_{2}\D t}\, K_{ij}^{n}W_{j}^{1,n}\nonumber \\
\delta x_{i}^{n+1}= & \delta x_{i}^{n}+\left[(1-w_{3}-w_{4}-w_{5})\left(\bar{H}_{ij}^{n}+\left(\partial_{j}\bar{H}_{ik}^{n}\right)\bar{x}_{k}^{n}\right)+w_{5}\left(\partial_{j}\bar{H}_{ik}^{n}\right)\bar{x}_{k}^{n+1}\right]\delta x_{j}^{n}\D t\label{eq:general_dx_np1}\\
+ & \left[w_{3}\left(\bar{H}_{ij}^{(p)}+\left(\partial_{j}\bar{H}_{ik}^{(p)}\right)\bar{x}_{k}^{(p)}\right)+w_{4}\left(\partial_{j}\bar{H}_{ik}^{(p)}\right)\bar{x}_{k}^{n+1}\right]\delta x_{j}^{(p)}\D t\nonumber \\
+ & \left(w_{4}\bar{H}_{ij}^{(p)}+w_{5}\bar{H}_{ij}^{n}\right)\delta x_{j}^{n+1}\D t\nonumber \\
+ & \begin{cases}
\left((1-w_{6})(\partial_{k}\bar{h}_{i}^{n})\delta x_{k}^{n}+w_{6}(\partial_{k}\bar{h}_{i}^{(p)})\delta x_{k}^{(p)}\right)\D t, & \mbox{or}\\
\partial_{k}\bar{h}_{i}^{(w)}\left((1-w_{6})\d x_{k}^{n}+w_{6}\d x_{k}^{(p)}\right)\D t
\end{cases}\nonumber \\
+ & \left((1-w_{7})\bar{K}_{ij}^{n}+w_{7}\bar{K}_{ij}^{(p)}\right)\left(\sqrt{w_{2}\D t}\, W_{j}^{1,n}+\sqrt{(1-w_{2})\D t}\, W_{j}^{2,n}\right),\nonumber 
\end{align}
where decorations and superscripts indicate where the derivatives
are evaluated, for example, $\partial_{k}\bar{h}_{i}^{(p)}=\partial_{k}h_{i}\left(\bar{\V x}^{(p)}\right)$,
and $\partial_{k}\bar{h}_{i}^{(w)}$ is evaluated at $(1-w_{6})\bar{\V x}^{n}+w_{6}\bar{\V x}^{(p)}$.
The analysis summarized in Appendix \ref{Appendix-Accuracy} shows
that this scheme is a second-order weakly accurate integrator for
(\ref{eq:dxbar_dt},\ref{eq:dx_dt_lin}) if the conditions (\ref{eq:second_order_cond})
hold.

This analysis leads us to the important conclusion that the same schemes
can be used not only to integrate time-dependent Langevin equations
(\ref{eq:TimeDepNoise}) but also to numerically linearize (\ref{eq:dx_dt_nonlin}),
and integrate the equations (\ref{eq:dxbar_dt},\ref{eq:dx_dt_lin})
to second order weakly. In particular, for fully explicit schemes
we recommend the midpoint scheme (\ref{eq:expl_midpoint}), and for
implicit schemes we recommend the implicit trapezoidal scheme (\ref{eq:impl_trapezoidal}),
with the identification $\M L\left(\V x\right)\equiv\M H\left(\V x\right)$
and $\V g(\V x)\equiv\V h(\V x)$. Depending on the interpretation
of the noise, the same integrators may also be first-order weak integrators
for the nonlinear equation (\ref{eq:dx_dt_nonlin}), in particular,
this is the case for Stratonovich noise. For kinetic noise, the RFD
term (\ref{eq:RFD_kinetic}) can be added to the corrector to obtain
the required stochastic drift terms.

\section{\label{sec:Overdamped}Fast-Slow Systems}

In this section we consider a generic finite-dimensional Langevin
equation of the form (\ref{eq:x_t_general}) in the case when there
are two variables, $\V x\leftarrow\left(\V x,\V y\right)$, where
$\V x\in\mathcal{R}^{N}$ is the slow (relevant) variable and $\V y\in\mathcal{R}^{M}$
is a fast variable. We focus on a rather general form of such a fast-slow
Langevin system,
\begin{align}
\left[\begin{array}{c}
\partial_{t}\V x\\
\partial_{t}\V y
\end{array}\right]= & -\left[\begin{array}{cc}
\M A & \epsilon^{-1}\M B\\
-\epsilon^{-1}\M B^{\star} & \epsilon^{-2}\M C
\end{array}\right]\left[\begin{array}{c}
\partial_{\V x}U\\
\partial_{\V y}U
\end{array}\right]+\sqrt{2k_{B}T}\,\left[\begin{array}{cc}
\M A_{\frac{1}{2}} & \M 0\\
\M 0 & \epsilon^{-1}\M C_{\frac{1}{2}}
\end{array}\right]\left[\begin{array}{c}
\V{\mathcal{W}}_{\V x}\left(t\right)\\
\V{\mathcal{W}}_{\V y}\left(t\right)
\end{array}\right]+\left(k_{B}T\right)\left[\begin{array}{c}
\partial_{\V x}\cdot\M A\\
-\epsilon^{-1}\partial_{\V x}\cdot\M B^{\star}
\end{array}\right]=\nonumber \\
= & -\M N\left(\V x\right)\left[\begin{array}{c}
\partial_{\V x}U\\
\partial_{\V y}U
\end{array}\right]+\left(2k_{B}T\right)^{\frac{1}{2}}\M M_{\frac{1}{2}}\left(\V x\right)\V{\mathcal{W}}(t)+\left(k_{B}T\right)\partial_{\V x}\cdot\M N\left(\V x\right),\label{eq:original_system}
\end{align}
where $\epsilon$ is a parameter that controls the separation of time
scales between the slow and fast variables, $\epsilon=1$ in the original
(inertial) dynamics. Here the linear operators $\M A\left(\V x\right)\succeq0$,
$\M B\left(\V x\right)$ and $\M C\left(\V x\right)\succeq0$ depend
\emph{only} on the slow variable $\V x$, and $\M A_{\frac{1}{2}}\left(\V x\right)$
and $\M C_{\frac{1}{2}}\left(\V x\right)$ satisfy the fluctuation-dissipation
balance condition (\ref{eq:FDB}), $\M A_{\frac{1}{2}}\M A_{\frac{1}{2}}^{\star}=\M A$
and $\M C_{\frac{1}{2}}\M C_{\frac{1}{2}}^{\star}=\M C$. The Langevin
equation with position-dependent friction (\ref{eq:inertial_Langevin})
is an example of this kind of system with the identification $\V y\equiv\sqrt{m}\V v$,
$U\left(\V x,\V y\right)=U\left(\V x\right)+y^{2}/2$, $\epsilon\equiv\sqrt{m}$,
$\M A=0$, $\M B=-\M I$, $\M C=\M{\gamma}$.

We will assume that the coarse-grained free energy $U\left(\V x,\V y\right)$
is \emph{separable} and \emph{quadratic} in the fast variable,
\begin{equation}
U\left(\V x,\V y\right)=U_{x}\left(\V x\right)+U_{y}\left(\V y\right)=U_{x}\left(\V x\right)+\frac{1}{2}\left(\V y-\bar{\V y}\left(\V x\right)\right)^{T}\M{\Psi}\left(\V x\right)\left(\V y-\bar{\V y}\left(\V x\right)\right),\label{eq:U_xy}
\end{equation}
so that the equilibrium distribution (invariant measure) of $\V y$
for \emph{fixed} $\V x$ is Gaussian with mean $\bar{\V y}\left(\V x\right)$
and covariance $\M{\Psi}\left(\V x\right)^{-1}$ that may depend on
the slow variable. In particular, in the model equation (\ref{eq:original_system})
the fast variable enters only linearly (but the slow variable enters
nonlinearly); this greatly simplifies the limiting dynamics \cite{AdiabaticElimination_1}
as $\epsilon\rightarrow0$. In the more general case one cannot write
a single SDE for the slow variable $\V x$ that describes all aspects
of the dynamics. Instead, depending on the type of observable and
the time scale one is interested in, different equations arise in
the limit.

As $\epsilon\rightarrow0$ there is an infinite separation of time
scales between $\V x$ and $\V y$, and we can perform adiabatic elimination
of the fast variable \cite{AdiabaticElimination_1}. The resulting
overdamped dynamics is expected to be a good approximation to the
original dynamics, $\epsilon=1$. In the limit $\epsilon\rightarrow0$,
it can be shown \cite{Averaging_Khasminskii,Averaging_Kurtz,ModeElimination_Papanicolaou}
(for a review see also the book \cite{AveragingHomogenization}) that
the \emph{limiting} or \emph{overdamped} dynamics for $\V x$ is
\begin{eqnarray}
\partial_{t}\V x & = & -\left(\M A+\M B\M C^{-1}\M B^{\star}\right)\partial_{\V x}F+\sqrt{2k_{B}T}\left(\M A_{\frac{1}{2}}\diamond\V{\mathcal{W}}_{\V x}+\M B\M C^{-1}\M C_{\frac{1}{2}}\diamond\V{\mathcal{W}}_{\V y}\right)\label{eq:limiting_equation}\\
 & = & -\left(\M A+\M B\M C^{-1}\M B^{\star}\right)\partial_{\V x}F+\sqrt{2k_{B}T}\left(\M A_{\frac{1}{2}}\V{\mathcal{W}}_{\V x}+\M B\M C^{-1}\M C_{\frac{1}{2}}\V{\mathcal{W}}_{\V y}\right)+\left(k_{B}T\right)\partial_{\V x}\cdot\left(\M A+\M B\M C^{-1}\M B^{\star}\right),\nonumber 
\end{eqnarray}
and is time-reversible with respect to the equilibrium distribution
$\sim\exp\left(-F\left(\V x\right)/k_{B}T\right)$. Here the symbol
$\diamond$ denotes the kinetic stochastic interpretation \cite{KineticStochasticIntegral_Ottinger}.
For future reference, we note that the last thermal drift in (\ref{eq:limiting_equation})
can be expanded using the chain rule,
\begin{equation}
\partial_{\V x}\cdot\left(\M B\M C^{-1}\M B^{\star}\right)=\partial_{\V x}\left(\M B\M C^{-1}\right):\M B^{\star}+\M B\M C^{-1}\left(\partial_{\V x}\cdot\M B^{\star}\right),\label{eq:BCB_drift}
\end{equation}
and similarly,
\[
\partial_{\V x}\cdot\M A=\partial_{\V x}\cdot\left(\M A_{\frac{1}{2}}\M A_{\frac{1}{2}}^{\star}\right)=\partial_{\V x}\left(\M A_{\frac{1}{2}}\right):\M A^{\star}+\M A_{\frac{1}{2}}\left(\partial_{\V x}\cdot\M A_{\frac{1}{2}}^{\star}\right),
\]
where the first term on the right hand side in index notation reads
$\left(\partial_{j}A_{ik}^{\myhalf}\right)A_{jk}^{\myhalf}$, and
is the only drift term that would have appeared if the overdamped
dynamics used a Stratonovich interpretation.

Here $F\left(\V x\right)$ is the free energy in the slow variables,
and can be obtained from the marginal distribution
\[
\exp\left(-\frac{F\left(\V x\right)}{k_{B}T}\right)\sim\exp\left(-\frac{U_{x}\left(\V x\right)}{k_{B}T}\right)\int\exp\left(-\frac{U_{y}\left(\V x\right)}{k_{B}T}\right)d\V y\sim\abs{\M{\Psi}\left(\V x\right)}^{-\frac{1}{2}}\exp\left(-\frac{U_{x}\left(\V x\right)}{k_{B}T}\right),
\]
giving the relationship
\[
F\left(\V x\right)=U_{x}\left(\V x\right)+\frac{k_{B}T}{2}\ln\abs{\M{\Psi}\left(\V x\right)}.
\]
Note that one can evaluate the derivative of the determinant using
Jacobi's formula,
\begin{equation}
\partial_{\V x}F=\partial_{\V x}U_{x}+\frac{k_{B}T}{2}\left(\M{\Psi}^{-1}:\partial_{\V x}\M{\Psi}\left(\V x\right)\right),\label{eq:dF_dx}
\end{equation}
where colon denotes a double contraction, $\left(\M{\Psi}^{-1}:\partial_{\V x}\M{\Psi}\right)_{k}=\Psi_{ij}^{-1}\partial_{k}\Psi_{ij}$.
One can construct a random finite difference approach (see Section
\ref{sub:RFD}) to evaluate this term in expectation, using the identity
\begin{equation}
\M{\Psi}^{-1}\left(\V x\right):\left(\partial_{\V x}\M{\Psi}\left(\V x\right)\right)=\lim_{\delta\rightarrow0}\frac{1}{\delta}\av{\left[\M{\Psi}^{-1}\left(\V x\right)\colon\left(\M{\Psi}\left(\V x+\delta\V W\right)-\M{\Psi}\left(\V x\right)\right)\right]\V W},\label{eq:determinant_RDF}
\end{equation}
where $\V W$ is a collection of i.i.d. standard normal variables.
This only requires a routine for evaluating the trace $\M{\Psi}^{-1}\colon\M{\Psi}$
once per time step, which is likely nontrivial in practice. However,
the RFD (\ref{eq:determinant_RDF}) avoids computing derivatives and
is certainly more efficient than computing $\M{\Psi}^{-1}:\partial_{\V x}\M{\Psi}$
using finite differences to evaluate the gradient $\partial_{\V x}\M{\Psi}$,
since this requires evaluating a trace for \emph{each} slow variable.
In this work we do not study models where $\M{\Psi}\left(\V x\right)$
depends on the slow variable further, and henceforth $F\left(\V x\right)\equiv U_{x}\left(\V x\right)$.

We now present two temporal integrators for solving (\ref{eq:limiting_equation}).
The first scheme can be seen as an application of the explicit midpoint
scheme (\ref{eq:expl_midpoint}) to the overdamped equation (\ref{eq:limiting_equation}),
and, aside from some of the stochastic drift terms, consists of applying
the explicit midpoint scheme to the original system (\ref{eq:original_system})
setting $\partial_{t}\V y=0$. Similarly, we also present an implicit
trapezoidal scheme (\ref{eq:impl_trapezoidal}) applied to the overdamped
equation (\ref{eq:limiting_equation}). In the explicit midpoint scheme
we use a random finite difference (RFD) approach to handle several
of the stochastic drift terms, as in (\ref{eq:RFD_Lang}), and in
the implicit trapezoidal scheme we show how one can use a Fixman-like
approach for one of the drift terms, as in (\ref{eq:Fixman_Lang}).

In the general nonlinear setting, the schemes presented in this section
are only first-order weakly accurate. They are deterministically second-order
accurate. With care, the schemes \emph{can }be second-order weak integrators
for the \emph{linearized} limiting dynamics, as we explained in Section
\ref{sec:Linearized}. Note that the thermal drift terms play no role
in the linearized overdamped dynamics because they are proportional
to the small noise variance, and only terms of order one half in the
noise variance are kept in the linearization. In order to achieve
second-order accuracy the schemes below require solving three linear
systems involving $\M C$ per time step. In cases when first-order
accuracy is sufficient, one can avoid the linear solve in the RFD
term; this can save significant computer time in cases when computing
the action of $\M C^{-1}$ is the dominant cost. Furthermore, in the
case of constant $\M C$, one can re-use the predictor's deterministic
increment in the corrector for a first order scheme requiring only
one linear solve. In this case, the role of the corrector stage is
simply to obtain all of the required stochastic drift terms \cite{BrownianBlobs}.

Of course, the trapezoidal and the midpoint schemes we presented here
do not exhaust all possibilities; they are representative of a broad
class of schemes that uses the original dynamics to approximate the
overdamped dynamics. By combining the techniques we described here
one can construct various schemes tailored to particular problems.
In each specific application, various terms may vanish, and a different
approach may be more efficient, more stable, or simpler to implement,
as we illustrate in Section \ref{sec:Applications} on several applications
in fluctuating hydrodynamics.

\subsection{Explicit Midpoint Scheme}

In the explicit midpoint scheme, the predictor is an Euler-Maruyama
step to the midpoint of the time step,
\begin{equation}
\left[\begin{array}{c}
\left(\V x^{p,n+\frac{1}{2}}-\V x^{n}\right)/\left(\D t/2\right)\\
\V 0
\end{array}\right]=-\left[\begin{array}{cc}
\M A & \M B\\
-\M B^{\star} & \M C
\end{array}\right]^{n}\left[\begin{array}{c}
\partial_{\V x}F\\
\M{\Psi}\left(\V x\right)\left(\V y-\bar{\V y}\left(\V x\right)\right)
\end{array}\right]^{n}+\sqrt{\frac{2k_{B}T}{\left(\D t/2\right)}}\,\left[\begin{array}{cc}
\M A_{\frac{1}{2}} & \M 0\\
\M 0 & \M C_{\frac{1}{2}}
\end{array}\right]^{n}\left[\begin{array}{c}
\V W_{\V x}^{n,1}\\
\V W_{\V y}^{n,1}
\end{array}\right],\label{eq:mid_pred}
\end{equation}
In terms of just $\V x$, the predictor step above is equivalent to
an Euler-Maruyama half-step for the limiting equation, without the
stochastic drift terms,
\begin{equation}
\V x^{p,n+\frac{1}{2}}=\V x^{n}-\frac{\D t}{2}\left(\M A+\M B\M C^{-1}\M B^{\star}\right)^{n}\left(\partial_{\V x}F\right)^{n}+\sqrt{k_{B}T\D t}\,\M A_{\frac{1}{2}}^{n}\V W_{\V x}^{n,1}+\sqrt{k_{B}T\D t}\,\left(\M B\M C^{-1}\M C_{\frac{1}{2}}\right)^{n}\V W_{\V y}^{n,1}.\label{eq:predictor_slow}
\end{equation}
In practical implementation, however, we first solve a linear system
for $\V y$ and then take an Euler-Maryama step for $\V x$ using
the obtained solution for $\V y$. This makes it very easy to convert
a semi-implicit code for simulating the original dynamics (\ref{eq:original_system})
to simulate the overdamped dynamics, using a much larger time step
size than possible for the original inertial dynamics. Note, however,
that a solver for linear systems involving $\M C$ must be implemented
and applied each time step; this will in general significantly increase
the cost of an overdamped step compared to a fully \emph{explicit}
scheme for the original dynamics. In practice, however, the extreme
stiffness in the original dynamics will force us to use a semi-implicit
scheme even for the original dynamics (for example, most fluid dynamics
codes treat viscosity, or, more generally, diffusion, semi-implicitly),
and the required linear solvers will already be available.

\subsubsection{First Order Method}

If first-order weak accuracy is sufficient, we can reuse the same
noise amplitude in the corrector stage as in the predictor stage,
and include RFD terms to capture the stochastic drift terms,

\begin{eqnarray}
\left[\begin{array}{c}
\left(\V x^{n+1}-\V x^{n}\right)/\D t\\
\V 0
\end{array}\right] & = & -\left[\begin{array}{cc}
\M A & \M B\\
-\M B^{\star} & \M C
\end{array}\right]^{p,n+\frac{1}{2}}\left[\begin{array}{c}
\partial_{\V x}F\\
\M{\Psi}\left(\V x\right)\left(\V y-\bar{\V y}\left(\V x\right)\right)
\end{array}\right]^{p,n+\frac{1}{2}}\label{eq:mid_corr_1st_order}\\
 & + & \sqrt{\frac{k_{B}T}{\D t}}\,\left[\begin{array}{cc}
\M A_{\frac{1}{2}} & \M 0\\
\M 0 & \M C_{\frac{1}{2}}
\end{array}\right]^{n}\left[\begin{array}{c}
\V W_{\V x}^{n,1}+\V W_{\V x}^{n,2}\\
\V W_{\V y}^{n,1}+\V W_{\V y}^{n,2}
\end{array}\right]\nonumber \\
 & + & \frac{k_{B}T}{\delta}\left[\begin{array}{c}
\M A\left(\V x^{n}+\delta\widetilde{\M W}^{n}\right)-\M A\left(\V x^{n}\right)\\
\M B^{\star}\left(\V x^{n}\right)-\M B^{\star}\left(\V x^{n}+\delta\widetilde{\M W}^{n}\right)
\end{array}\right]\widetilde{\M W}^{n}.\nonumber 
\end{eqnarray}
Note that here we can re-use the same random numbers $\widetilde{\M W}^{n}$
for both stochastic drift terms, but this is not necessary.

The corrector step includes random finite differences to capture the
stochastic drift terms $\left(k_{B}T\right)\partial_{\V x}\cdot\M A$
and $\left(k_{B}T\right)\M B\M C^{-1}\left(\partial_{\V x}\cdot\M B^{\star}\right)$,
as we did in (\ref{eq:RFD_Lang}). The remaining drift term $\left(k_{B}T\right)\partial_{\V x}\left(\M B\M C^{-1}\right):\M B^{\star}$
is obtained from the predictor step in the spirit of Runge-Kutta schemes,
as can be confimed by a Taylor series analysis (see Appendix \ref{sec:Appendix-Kinetic}
for further details). To see this, note that the stochastic increment
in $\V x^{p,n+1}$ involving $\V W_{\V y}^{n,1}$ is
\[
\D{\V x}^{p}=\sqrt{\D t\, k_{B}T}\,\left(\M B\M C^{-1}\M C_{\frac{1}{2}}\right)^{n}\V W_{\V y}^{n,1}.
\]
In the corrector, we have the stochastic increment
\[
\D{\V x}^{c}=\left(\M B\M C^{-1}\right)^{p,n+\frac{1}{2}}\sqrt{\D{t\,}k_{B}T}\,\M C_{\frac{1}{2}}^{n}\left(\V W_{\V y}^{n,1}+\V W_{y}^{n,2}\right).
\]
If we expand $\left(\M B\M C^{-1}\right)^{p,n+1}$ to first-order
around $\V x^{n}$, we see that $\D{\V x}^{c}$ contains a term $\sim\partial_{\V x}\left(\M B\M C^{-1}\right)^{n}\D{\V x}^{p}\left(\M C_{\frac{1}{2}}^{n}\V W_{\V y}^{n,1}\right)$,
more precisely, in index notation, the $i$-th component of the additional
term is
\[
\sqrt{\D t\, k_{B}T}\partial_{l}\left(\M B\M C^{-1}\right)_{ij}\D x_{l}^{p}C_{jk}^{\myhalf}W_{k}^{y,1}=\left(k_{B}T\right)\D t\;\partial_{l}\left(\M B\M C^{-1}\right)_{ij}\; B_{lm}C_{mn}^{-1}C_{np}^{\myhalf}C_{jk}^{\myhalf}\; W_{k}^{y,1}W_{p}^{y,1},
\]
evaluated at time step $n$, where $\partial_{k}\equiv\partial/\partial x_{k}$.
Upon taking expectation values, $\av{W_{k}^{y,1}W_{p}^{y,1}}=\delta_{k,p}$,
we obtain the required drift term $\partial_{\V x}\left(\M B\M C^{-1}\right):\M B^{\star}$
in (\ref{eq:BCB_drift}), since
\begin{equation}
\partial_{l}\left(\M B\M C^{-1}\right)_{ij}\, B_{lm}\, C_{mn}^{-1}C_{nk}^{\myhalf}\left(C^{\myhalf}\right)_{jk}^{\star}=\partial_{l}\left(\M B\M C^{-1}\right)_{ij}\, B_{lm}\,\delta_{jm}=\partial_{k}\left(\M B\M C^{-1}\right)_{ij}\, B_{kj}.\label{eq:BCB_drift_term}
\end{equation}

\subsubsection{Second Order Method}

If we want to obtain second-order weak accuracy for the \emph{linearized}
overdamped dynamics, we should evaluate the noise in the corrector
at the predicted midpoint value, as in the explicit midpoint algorithm
(\ref{eq:expl_midpoint}). This is however only consistent with a
Stratonovich interpretation of the noise in the overdamped dynamics
and is not consistent with the kinetic interpretation we seek. In
order to be consistent with a kinetic interpretation, we need to add
RFD terms to capture the correct stochastic drift terms (see Appendix
\ref{sec:Appendix-Kinetic}),
\begin{eqnarray}
\left[\begin{array}{c}
\left(\V x^{n+1}-\V x^{n}\right)/\D t\\
\V 0
\end{array}\right] & = & -\left[\begin{array}{cc}
\M A & \M B\\
-\M B^{\star} & \M C
\end{array}\right]^{p,n+\frac{1}{2}}\left[\begin{array}{c}
\partial_{\V x}F\\
\M{\Psi}\left(\V x\right)\left(\V y-\bar{\V y}\left(\V x\right)\right)
\end{array}\right]^{p,n+\frac{1}{2}}\label{eq:mid_corr}\\
 & + & \sqrt{\frac{k_{B}T}{\D t}}\,\left[\begin{array}{cc}
\M A_{\frac{1}{2}}^{p,n+\frac{1}{2}} & \M 0\\
\M 0 & \M C_{\frac{1}{2}}^{p,n+\frac{1}{2}}
\end{array}\right]\left[\begin{array}{c}
\V W_{\V x}^{n,1}+\V W_{\V x}^{n,2}\\
\V W_{\V y}^{n,1}+\V W_{\V y}^{n,2}
\end{array}\right]\nonumber \\
 & + & \frac{k_{B}T}{\delta}\left[\begin{array}{c}
\M A_{\frac{1}{2}}^{n}\left(\M A_{\frac{1}{2}}^{\star}\left(\V x^{n}+\delta\widetilde{\V W}_{A}^{n}\right)-\M A_{\frac{1}{2}}^{\star}\left(\V x^{n}\right)\right)\\
\M B^{\star}\left(\V x^{n}+\delta\widetilde{\V W}_{A}^{n}\right)-\M B^{\star}\left(\V x^{n}\right)
\end{array}\right]\widetilde{\V W}_{A}^{n}\nonumber \\
 & - & \frac{k_{B}T}{\delta}\left[\begin{array}{c}
0\\
\M C_{\frac{1}{2}}\left(\V x^{n}+\delta\left(\M B\M C^{-1}\M C_{\frac{1}{2}}\right)^{n}\widetilde{\V W}_{C}^{n}\right)-\M C_{\frac{1}{2}}\left(\V x^{n}\right)
\end{array}\right]\widetilde{\V W}_{C}^{n},\nonumber 
\end{eqnarray}
where $\widetilde{\V W}_{A}^{n}$ and $\widetilde{\V W}_{C}^{n}$
are independently-generated random vectors. As we show in Appendix
\ref{sec:Appendix-Kinetic}, the scheme (\ref{eq:mid_corr}) is weakly
first-order accurate in general, while also achieving second order
weak accuracy for the linearized overdamped dynamics. In \cite{BrownianBlobs}
we successfully used the scheme (\ref{eq:mid_pred},\ref{eq:mid_corr})
to integrate the equations of Brownian Dynamics, which result when
one eliminates the fast velocity degrees of freedom from a system
of equations for the motion of particles immersed in a fluctuating
Stokes fluid.

The midpoint scheme is (\ref{eq:mid_pred},\ref{eq:mid_corr}) a \emph{second-order}
weak integrator for the \emph{linearized} overdamped equations. This
is because it can be seen as an application of the explicit midpoint
scheme (\ref{eq:expl_midpoint}) to the limiting dynamics (\ref{eq:limiting_equation}),
which we concluded in Section \ref{sub:Linearization} to be a second-order
integrator for linearized Langevin equations. This shows the importance
of carefully selecting where to evaluate the noise amplitude in the
corrector stage in the nonlinear setting, and balancing this with
RFD terms to ensure consistency with the nonlinear equations. In the
case of constant $\M C$, one can omit the last line in (\ref{eq:mid_corr}),
similarly, if $\M A$ is constant one can omit the corresponding RFD
term. Furthermore, in some cases, such as the specific example studied
in Section \ref{sub:ConstCoeff}, there is no difference between a
Stratonovich or a kinetic interpretation of the limiting dynamics
and one can use the corrector (\ref{eq:mid_corr}) without the RFD
terms on the last two lines of (\ref{eq:mid_corr}).

\subsection{Implicit Trapezoidal Scheme}

In this section we explain how the implicit trapezoidal scheme (\ref{eq:impl_trapezoidal})
can be used to simulate the overdamped dynamics (\ref{eq:limiting_equation}).
We will assume that
\[
\M A\left(\V x\right)\,\partial_{\V x}F\left(\V x\right)\equiv\M L\left(\V x\right)\V x
\]
and treat this term semi-implicitly. All remaining terms, including
the stochastic drift term $\left(\M B\M C^{-1}\M B^{\star}\right)\partial_{\V x}F$,
which arises due to the elimination of the fast variable, will be
handled explicitly.

The predictor step consists of taking an overdamped step for $\V y$,
which simply amounts to deleting the term $\partial_{t}\V y$ in (\ref{eq:original_system}),
followed by an implicit trapezoidal step for $\V x$. Symbolically,
\begin{equation}
\left[\begin{array}{c}
\left(\V x^{p,n+1}-\V x^{n}\right)/\D t\\
\V 0
\end{array}\right]=-\left[\begin{array}{c}
\frac{1}{2}\M L^{n}\left(\V x^{n}+\V x^{p,n+1}\right)+\left(\M B\M{\Psi}\right)^{n}\left(\V y^{n}-\bar{\V y}^{n}\right)\\
-\left(\M B^{\star}\partial_{\V x}F\right)^{n}+\left(\M C\M{\Psi}\right)^{n}\left(\V y^{n}-\bar{\V y}^{n}\right)
\end{array}\right]+\sqrt{\frac{2k_{B}T}{\D t}}\,\left[\begin{array}{cc}
\M A_{\frac{1}{2}} & \M 0\\
\M 0 & \M C_{\frac{1}{2}}
\end{array}\right]^{n}\left[\begin{array}{c}
\V W_{\V x}^{n}\\
\V W_{\V y}^{n}
\end{array}\right].\label{eq:trap_pred}
\end{equation}
Note that here we have omitted all thermal drift terms; we will rely
on the corrector to obtain those. The corrector step consists of solving
the following linear system for $\V x^{n+1}$ and $\V y^{p,n+1}$,
\begin{eqnarray}
\left[\begin{array}{c}
\left(\V x^{n+1}-\V x^{n}\right)/\D t\\
\V 0
\end{array}\right] & = & \left[\begin{array}{c}
\D{\V x}/\D t\\
\D{\V y}/\D t
\end{array}\right]\label{eq:trap_corr}\\
 & - & \left[\begin{array}{c}
\frac{1}{2}\left(\M L^{n}\V x^{n}+\M L^{p,n+1}\V x^{n+1}\right)+\frac{1}{2}\left(\M B\M{\Psi}\right)^{p,n+1}\left(\V y^{p,n+1}-\bar{\V y}^{p,n+1}\right)+\frac{1}{2}\left(\M B\M{\Psi}\right)^{n}\left(\V y^{n}-\bar{\V y}^{n}\right)\\
-\left(\M B^{\star}\partial_{\V x}F\right)^{p,n+1}+\left(\M C\M{\Psi}\right)^{p,n+1}\left(\V y^{p,n+1}-\bar{\V y}^{p,n+1}\right)
\end{array}\right].\nonumber 
\end{eqnarray}
where the stochastic increments $\D{\V x}$ and $\D{\V y}$ are given
in (\ref{eq:stoch_incr}).

The stochastic increments $\D{\V x}$ and $\D{\V y}$ need to be carefully
constructed in order to obtain the correct drift terms, and can be
approximated in one of two ways. For the case when $\M A$ is invertible,
we can use a Fixman like approach to obtain the drift term $\left(k_{B}T\right)\partial_{\V x}\cdot\M A$
in the corrector step, just as we illustrated for the simple Langevin
equation in (\ref{eq:Fixman_Lang}), 
\begin{eqnarray}
\left[\begin{array}{c}
\D{\V x}/\D t\\
\D{\V y}/\D t
\end{array}\right] & = & \sqrt{\frac{2k_{B}T}{\D t}}\,\left[\begin{array}{cc}
\frac{1}{2}\left(\M A^{n}+\M A^{p,n+1}\right)\left(\M A^{-1}\M A_{\frac{1}{2}}\right)^{n} & \M 0\\
\M 0 & \M C_{\frac{1}{2}}^{n}
\end{array}\right]\left[\begin{array}{c}
\V W_{\V x}^{n}\\
\V W_{\V y}^{n}
\end{array}\right]\label{eq:stoch_incr}\\
 & + & 2\frac{k_{B}T}{\delta}\left[\begin{array}{c}
\V 0\\
\M B^{\star}\left(\V x^{n}\right)-\M B^{\star}\left(\V x^{n}+\delta\widetilde{\M W}^{n}\right)
\end{array}\right]\widetilde{\M W}^{n}.\nonumber 
\end{eqnarray}
The scheme (\ref{eq:trap_corr},\ref{eq:stoch_incr}) is only first-order
weakly accurate even for the linearized overdamped dynamics. 

If we want to obtain second-order weak accuracy for the \emph{linearized}
overdamped dynamics, we should evaluate the noise in the corrector
at the predicted value, as in the implicit trapezoidal algorithm (\ref{eq:impl_trapezoidal}).
In this case we need to capture the remaining terms with an RFD approach,
as we did in (\ref{eq:mid_corr}),
\begin{align}
\left[\begin{array}{c}
\D{\V x}/\D t\\
\D{\V y}/\D t
\end{array}\right]= & \sqrt{\frac{2k_{B}T}{\D t}}\,\left[\begin{array}{cc}
\frac{1}{2}\left(\M A_{\frac{1}{2}}^{n}+\M A_{\frac{1}{2}}^{p,n+1}\right) & \M 0\\
\M 0 & \M C_{\frac{1}{2}}^{p,n+1}
\end{array}\right]\left[\begin{array}{c}
\V W_{\V x}^{n}\\
\V W_{\V y}^{n}
\end{array}\right]\label{eq:trap_stoch_incr_2nd}\\
+ & \frac{k_{B}T}{\delta}\left[\begin{array}{c}
\M A_{\frac{1}{2}}^{n}\left(\M A_{\frac{1}{2}}^{\star}\left(\V x^{n}+\delta\widetilde{\V W}_{A}^{n}\right)-\M A_{\frac{1}{2}}^{\star}\left(\V x^{n}\right)\right)\\
2\left(\M B^{\star}\left(\V x^{n}+\delta\widetilde{\V W}_{A}^{n}\right)-\M B^{\star}\left(\V x^{n}\right)\right)
\end{array}\right]\widetilde{\V W}_{A}^{n}\nonumber \\
- & \frac{k_{B}T}{\delta}\left[\begin{array}{c}
0\\
\left(2\M C_{\frac{1}{2}}\left(\V x^{n}+\delta\left(\M B\M C^{-1}\M C_{\frac{1}{2}}\right)^{n}\widetilde{\V W}_{C}^{n}\right)-\M C_{\frac{1}{2}}\left(\V x^{n}\right)\right)
\end{array}\right]\widetilde{\V W}_{C}^{n}.\nonumber 
\end{align}
Note that the computation of $\left(\M B\M C^{-1}\M C_{\frac{1}{2}}\right)^{n}\widetilde{\V W}_{C}^{n}$
involves solving a linear system involving $\M C$ and will thus,
generally, significantly increase the computational effort per time
step. In the case of constant $\M A$ or $\M C$, one can omit the
corresponding RFD terms to simplify the scheme. For a Stratonovich
interpretation of the limiting dynamics, one simply omits the last
two terms of (\ref{eq:trap_stoch_incr_2nd}).

\section{\label{sub:ConstCoeff}\label{sec:Applications}Fluctuating Hydrodynamics:
Tracer Diffusion}

In this section we apply the techniques we developed in this work
to the fluctuating hydrodynamics problems described in Section \ref{sub:FHD}.
In the example considered here many of the stochastic drift terms
present in the more general case are not present. In \cite{BrownianBlobs}
we presented a numerical method for performing Brownian dynamics for
particles suspended in a fluid, based on treating the fluid velocity
as a fast degree of freedom compared to the positions of the particles.
That example includes the majority of the stochastic drift terms that
appear in a general setting, except that $\M C$ is constant, and
employs our midpoint scheme for overdamped dynamics essentially in
its full generality.

We model the diffusion of a concentration field that is passively
advected by the randomly fluctuating fluid velocity, as we first discussed
in Section \ref{sub:FHD}. In the nonlinear overdamped setting, the
methods presented here can be used to model diffusion of labeled or
tracer particles in liquids over a broad range of length scales, as
we did in Ref. \cite{DiffusionJSTAT}. In the linearized setting,
the same methods can be used to study the spatio-temporal spectrum
of Gaussian fluctuations around steady or dynamic deterministic flows
\cite{FluctHydroNonEq_Book}, as we first did in Ref. \cite{LLNS_Staggered}
for a steady state and extend in this work to a dynamic setting in
Section \ref{sec:GRADFLEX}. In Section \ref{sec:SORET} we present
an application of the methods developed here to study the dynamic
structure factors in binary fluid mixtures subjected to a small temperature
gradient, in the presence of gravity and confinement \cite{SoretDiffusion_Croccolo}.

We consider the velocity-concentration system (\ref{eq:fluct_NS},\ref{eq:c_eq_original})
in a simplified setting in which the noise in the concentration equation
is additive and we omit the nonlinear term $\V v\cdot\grad\V v$,
\begin{eqnarray}
\partial_{t}\V v & = & \M{\mathcal{P}}\left(\nu\grad^{2}\V v+\sqrt{2\rho^{-1}\nu k_{B}T}\,\grad\cdot\M{\mathcal{W}}-\beta c\,\V g\right),\label{eq:const-coeff}\\
\partial_{t}c & = & -\grad\cdot\left(c\V v\right)+\chi_{0}\grad^{2}c+\grad\cdot\left(\sqrt{2\chi_{0}\vartheta_{0}}\,\M{\mathcal{W}}_{c}\right),\nonumber 
\end{eqnarray}
where $\nu=\eta/\rho$ and $\vartheta_{0}=\rho^{-1}mc_{0}$, where
$m$ is the mass of the tracer particles and $c_{0}$ is a reference
concentration. For simplicity we have replaced $\V u=\M{\sigma}\star\V v$
by $\V v$, assuming that the filtering is done by an implicit truncation
of the SPDE at small scales; this is naturally performed in the finite-volume
discretizations of this system described in \cite{DFDB}. Note that
for incompressible $\V v$ we have $\grad\cdot\left(c\V v\right)=\V v\cdot\grad c$.
In (\ref{eq:const-coeff}) the constraint $\grad\cdot\V v=0$ is enforced
by the Helmholtz projection operator $\M{\mathcal{P}}=\M I-\M{\mathcal{G}}\left(\M{\mathcal{D}}\M{\mathcal{G}}\right)^{-1}\M{\mathcal{D}}$,
where $\M{\mathcal{D}}\equiv\grad\cdot$ denotes the divergence operator
and $\M{\mathcal{G}}\equiv\grad$ the gradient operator with the appropriate
boundary conditions taken into account.

The coupled velocity-concentration system (\ref{eq:const-coeff})
can be written as an infinite-dimensional system of the form (\ref{eq:original_system})
with the identification $\V v\equiv\V y$ as the (potentially) fast
variable and $c\equiv\V x$ as the slow variable. We have a quadratic
(Gaussian) coarse-grained energy \emph{functional }that includes a
contribution due to the (excess) gravitational potential energy of
the tracer particles of excess mass (over the solvent) of $\beta m$,
\begin{equation}
U\left[\V v\left(\cdot\right),\, c\left(\cdot\right)\right]=\frac{\rho}{2}\int v^{2}\left(\V r\right)\, d\V r+\beta\rho\int c\left(\V r\right)\left(\V r\cdot\V g\right)\, d\V r+\frac{k_{B}T}{2\vartheta_{0}}\int c^{2}\left(\V r\right)\, d\V r,\label{eq:U_FNS-2}
\end{equation}
which corresponds to $\V{\bar{y}\equiv\bar{v}}=\V 0$, $\M{\Psi}\left(\V x\right)\equiv\rho\M I/2$
a multiple of the identity, and quadratic $U\left(\V x\right)\equiv\left(2\vartheta_{0}\right)^{-1}k_{B}T\int c^{2}\left(\V r\right)\, d\V r$.
The mobility %
\footnote{Note that the advective part of the mobility operator we use here
is slightly different from that in \cite{DFDB} because here we use
the conservative $\grad\cdot\left(c\V v\right)$ rather than the advective
form $\V v\cdot\grad c$, as required for momentum conservation in
the presence of gravity.%
} and noise operators can be taken to be
\[
\M N\left[c\left(\cdot\right)\right]=\left[\begin{array}{cc}
\M A & \M B\\
-\M B^{\star} & \M C
\end{array}\right]\equiv-\left[\begin{array}{cc}
\left(k_{B}T\right)^{-1}\chi_{0}\vartheta_{0}\grad^{2} & -\rho^{-1}\grad\cdot c\M{\mathcal{P}}\\
-\rho^{-1}\M{\mathcal{P}}c\grad & \rho^{-1}\nu\left(\M{\mathcal{P}}\grad^{2}\M{\mathcal{P}}\right)
\end{array}\right]
\]
and
\[
\M M_{\myhalf}\left[c\left(\cdot\right)\right]=\left[\begin{array}{cc}
\M A_{\frac{1}{2}} & \M 0\\
\M 0 & \epsilon^{-1}\M C_{\frac{1}{2}}
\end{array}\right]=\left[\begin{array}{cc}
\sqrt{\chi_{0}\vartheta_{0}/k_{B}T}\,\grad\cdot & \M 0\\
\M 0 & \sqrt{\rho^{-1}\nu}\,\M{\mathcal{P}}\grad\cdot
\end{array}\right],
\]
where differential operators act on everything to their right and
we note that only the operator $\M B\left[c\left(\cdot\right)\right]$
is a functional of the slow variable $c$. The diagonal blocks of
the mobility operator $\M N$ generate momentum and (bare) mass diffusion
and the corresponding noise terms. The upper right block of $\M N$
generates the advective term $-\V v\cdot\grad c$ in the concentration
equation, while the lower left block of $\M N$ generates the gravity
term in the velocity equation,
\[
-\rho^{-1}\M{\mathcal{P}}c\grad\left(\rho\beta\left(\V r\cdot\V g\right)+\frac{k_{B}T}{2\vartheta_{0}}c\right)=-\beta\M{\mathcal{P}}c\,\V g-\rho^{-1}\M{\mathcal{P}}\grad\left(\frac{c^{2}}{2}\right)=-\beta\M{\mathcal{P}}c\,\V g,
\]
where we used the fact that projection eliminates pure gradients.
This last property allows for a key simplification, namely, the stochastic
drift term involving $\partial_{\V x}\cdot\M B^{\star}\equiv\partial_{\V c}\cdot\M B^{\star}\left[c\left(\cdot\right)\right]$
can be omitted, and there is no difference between a kinetic and a
Stratonovich interpretation of the overdamped dynamics \cite{DiffusionJSTAT}.
We take advantage of these properties in the algorithms presented
next. It is important to note that these simplifying properties are
also valid after (careful) spatial discretization of the SPDEs \cite{DFDB}.
Also note that spatially-discretized white noise acquires an additional
factor of $\D V^{-\myhalf}$, where $\D V$ is the volume of the grid
cells \cite{DFDB}.

Note that a more complete but formal calculation \cite{DDFT_Hydro}
using the true ideal gas entropy functional (\ref{eq:U_FNS}) would
set $\M A=\chi_{0}m/\left(\rho k_{B}T\right)\,\grad\cdot c\grad$,
where all differential operators act to their right, and obtain multiplicative
noise as well as an additional \emph{barodiffusion} term in the concentration
equation,
\begin{equation}
\partial_{t}c=-\grad\cdot\left(c\V v\right)+\grad\cdot\left[\chi_{0}\left(\grad c+\frac{\beta m\V g}{k_{B}T}c\right)\right]+\grad\cdot\left(\sqrt{2\chi_{0}\rho^{-1}mc}\,\M{\mathcal{W}}_{c}\right).\label{eq:c_eq_barodiff}
\end{equation}
The barodiffusion term is only important in the presence of very high
gravitational energies such as in ultracentrifuges, and we will neglect
it here and use (\ref{eq:c_eq_original}) instead.

\subsection{Inertial Equations}

For the full inertial dynamics (\ref{eq:const-coeff}), applying our
predictor-corrector methods would require treating the diffusion of
both momentum and mass with the same scheme, i.e., both would need
to treated implicitly or both treated explicitly. Similarly, if a
semi-implicit method is used, two linear solves would be required,
one for the predictor and one for the corrector stage. However, we
can take specific advantage of the structure of (\ref{eq:const-coeff})
and use a \emph{split} approach, in which we handle concentration
and velocity differently, for example, we can treat viscosity implicitly
(since velocity is a faster variable) and treat mass diffusion explicitly
(since concentration is a slower variable). In Algorithm \ref{alg:InertialConstant}
{} we present an optimized split scheme for integrating (\ref{eq:const-coeff}),
which only requires a single fluid solve per time step. Note that
this scheme can be made second-order accurate deterministically for
the fully nonlinear Navier-Stokes equation by using a time-lagged
(multi-step) scheme for the advective term $\rho\V v\cdot\grad\V v=\rho\grad\cdot\left(\V v\otimes\V v\right)$.
Here we treat concentration implicitly but an explicit treatment is
also possible. In Section \ref{sec:SORET} we use Algorithm \ref{alg:InertialConstant}
to study the difference between the inertial and overdamped dynamics
for a problem involving a fluid sample subjected to a concentration
gradient, in the presence of gravity.

\begin{algorithm}
\caption{\label{alg:InertialConstant}Split integrator for the inertial dynamics
(\ref{eq:const-coeff}), as implemented in the IBAMR software framework
\cite{IBAMR}.}

\begin{enumerate}
\item In the predictor step for concentration, solve for $c^{p,n+1}$, 
\[
\frac{c^{p,n+1}-c^{n}}{\D t}=-\V v^{n}\cdot\grad c^{n}+\chi_{0}\grad^{2}\left(\frac{c^{n}+c^{p,n+1}}{2}\right)+\grad\cdot\left(\sqrt{\frac{2\chi_{0}\vartheta_{0}}{\D t\D V}}\,\V W_{c}^{n}\right).
\]

\item Solve the time-dependent Stokes or Navier-Stokes system for $\V v^{n+1}$
and $\pi^{n+\frac{1}{2}}$,
\begin{align*}
\rho\frac{\V v^{n+1}-\V v^{n}}{\D t}+\rho\left(\V v\cdot\grad\V v\right)^{n+\myhalf}+\grad\pi^{n+\frac{1}{2}} & =\eta\grad^{2}\left(\frac{\V v^{n+1}+\V v^{n}}{2}\right)+\grad\cdot\left(\sqrt{\frac{2\eta\, k_{B}T}{\D t\D V}}\,\M W^{n}\right)-\rho\beta\left(\frac{c^{n}+c^{p,n+1}}{2}\right)\V g\\
\grad\cdot\V v^{n+1} & =0.
\end{align*}
The nonlinear advective term $\left(\V v\cdot\grad\V v\right)^{n+\myhalf}$
can be omitted in the Stokes limit, or approximated to second-order
accuracy using an Adams-Bashforth approach.
\item Correct the concentration by solving
\[
\frac{c^{n+1}-c^{n}}{\D t}=-\left(\frac{\V v^{n+1}+\V v^{n}}{2}\right)\cdot\grad\left(\frac{c^{n}+c^{p,n+1}}{2}\right)+\chi_{0}\grad^{2}\left(\frac{c^{n}+c^{n+1}}{2}\right)+\grad\cdot\left(\sqrt{\frac{2\chi_{0}\vartheta_{0}}{\D t\D V}}\,\V W_{c}^{n}\right).
\]
\end{enumerate}
\end{algorithm}

The analysis presented in this work does not directly apply to split
schemes and confirming second-order weak accuracy requires custom
analysis. Empirical order of accuracy tests can also be used but note
that these can be misleading since error terms that dominate for very
small time step sizes may be negligible for time step sizes of interests.
For very small $\D t$, statistical errors are often much larger than
the truncation errors, making empirical convergence studies computationally
infeasible.

For completeness, in Algorithm \ref{alg:VarCoeffInert} we apply our
implicit trapezoidal integrator (\ref{eq:impl_trapezoidal}) to the
variable-coefficient inertial equations (\ref{eq:variable_coeff});
this integrator requires two concentration and two velocity (Stokes)
linear solves per time step. Our analysis shows that this is a second-order
weak integrator for the \emph{linearized} inertial equations. We do
not use this integrator in this work because a constant-coefficient
incompressible approximation is appropriate in the (passive tracer)
applications we study here. Note that for incompressible $\V v$ the
conservative and advective forms are equivalent, $\grad\cdot\left(c\V v\right)=\V v\cdot\grad c$;
in the numerical schemes it is preferred to use the conservative form
to ensure strict conservation of mass and momentum even when imposing
the divergence-free constraint on the velocity only to some finite
threshold. Note that the nonlinear advective terms can alternatively
be handled using a midpoint rule following (\ref{eq:impl_trap_expl_mid}).
For example, $\left[\left(c\V v\right)^{n}+\left(c\V v\right)^{p,n+1}\right]/2$
can be replaced by 
\[
\left(\frac{c^{n}+c^{p,n+1}}{2}\right)\left(\frac{\V v^{n}+\V v{}^{p,n+1}}{2}\right),
\]
without affecting the order of accuracy.

\begin{algorithm}
\caption{\label{alg:VarCoeffInert}Unsplit implicit trapezoidal temporal integrator
for the variable-coefficient inertial equations (\ref{eq:variable_coeff}).}

\begin{enumerate}
\item In the predictor step for concentration, solve for $c^{p,n+1}$,
\[
\frac{c^{p,n+1}-c^{n}}{\D t}=-\grad\cdot\left(c\V v\right)^{n}+\frac{1}{2}\grad\cdot\left[\chi^{n}\left(\grad c^{n}+\grad c^{p,n+1}\right)\right]+\grad\cdot\left(\sqrt{\frac{2\left(\chi\rho^{-1}\mu_{c}^{-1}\right)^{n}k_{B}T}{\D t\D V}}\,\V W_{c}^{n}\right),
\]
and solve (independently) the variable-coefficient Stokes system for
$\V v^{p,n+1}$ and $\pi^{p,n+\frac{1}{2}}$ \cite{StokesKrylov},
\begin{eqnarray*}
\rho\frac{\V v^{p,n+1}-\V v^{n}}{\D t}+\grad\pi^{p,n+\frac{1}{2}} & = & -\rho\grad\cdot\left(\V v\otimes\V v\right)^{n}-\beta\rho c^{n}\,\V g\\
 & + & \grad\cdot\left[\frac{\eta^{n}}{2}\left(\bar{\grad}\V v^{n}+\bar{\grad}\V v^{p,n+1}\right)+\sqrt{\frac{2\eta^{n}k_{B}T}{\D t\,\D V}}\,\M W^{n}\right]\\
\grad\cdot\V v^{p,n+1} & = & 0.
\end{eqnarray*}

\item Correct the concentration by solving,
\begin{eqnarray*}
\frac{c^{n+1}-c^{n}}{\D t} & = & -\frac{1}{2}\grad\cdot\left[\left(c\V v\right)^{n}+\left(c\V v\right)^{p,n+1}\right]+\frac{1}{2}\grad\cdot\left[\chi^{n}\grad c^{n}+\chi^{p,n+1}\grad c^{n+1}\right]\\
 & + & \frac{1}{2}\grad\cdot\left[\left(\sqrt{\frac{2\left(\chi\rho^{-1}\mu_{c}^{-1}\right)^{n}k_{B}T}{\D t\D V}}+\sqrt{\frac{2\left(\chi\rho^{-1}\mu_{c}^{-1}\right)^{p,n+1}k_{B}T}{\D t\D V}}\right)\V W_{c}^{n}\right],
\end{eqnarray*}
and correct the velocity by solving (independently) the the variable-coefficient
Stokes system for $\V v^{n+1}$ and $\pi^{n+\frac{1}{2}}$ \cite{StokesKrylov},
\begin{eqnarray*}
\rho\frac{\V v^{n+1}-\V v^{n}}{\D t}+\grad\pi^{n+\frac{1}{2}} & = & -\frac{\rho}{2}\grad\cdot\left[\left(\V v\otimes\V v\right)^{n}+\left(\V v\otimes\V v\right)^{p,n+1}\right]-\frac{\beta\rho}{2}\left(c^{n}+c^{p,n+1}\right)\V g\\
 & + & \frac{1}{2}\grad\cdot\left(\eta^{n}\bar{\grad}\V v^{n}+\eta^{p,n+1}\bar{\grad}\V v^{n+1}\right)+\\
 & + & \frac{1}{2}\grad\cdot\left[\left(\sqrt{\frac{2\eta^{n}k_{B}T}{\D t\,\D V}}+\sqrt{\frac{2\eta^{p,n+1}k_{B}T}{\D t\,\D V}}\right)\,\M W^{n}\right]\\
\grad\cdot\V v^{n+1} & = & 0.
\end{eqnarray*}
\end{enumerate}
\end{algorithm}

\subsection{Overdamped Equations}

The overdamped limit of (\ref{eq:const-coeff}) is the Stratonovich
SPDE \cite{DiffusionJSTAT}
\begin{eqnarray}
\partial_{t}c & = & \beta\rho\eta^{-1}\left(\M G\star c\V g\right)\cdot\grad c+\sqrt{2\eta^{-1}k_{B}T}\,\left(\M G_{\myhalf}\M{\mathcal{W}}\right)\odot\grad c\label{eq:limiting_Strato_const}\\
 & + & \chi_{0}\grad^{2}c+\grad\cdot\left(\sqrt{2\chi_{0}\vartheta_{0}}\,\M{\mathcal{W}}_{c}\right),\nonumber 
\end{eqnarray}
where the notation is explained in Section \ref{sub:OverdampedFHD}.
In Algorithm \ref{alg:OverdampedConstant}%
{} we give a temporal integrator for this equation, which we first proposed
in Appendix B of Ref. \cite{DiffusionJSTAT}. This algorithm can be
seen as an application of the implicit trapezoidal method (\ref{eq:impl_trap_expl_mid})
to (\ref{eq:limiting_Strato_const}), with the identifications
\[
\M L\equiv\chi_{0}\grad^{2},\quad\mbox{and}\quad\V g\left[c\left(\cdot\right)\right]=\beta\rho\eta^{-1}\left(\M G\star c\V g\right)\cdot\grad c,
\]
and
\[
\M K\left[c\left(\cdot\right)\right]\circ\V{\mathcal{W}}\left(t\right)\equiv\sqrt{2\eta^{-1}k_{B}T}\,\left(\M G_{\myhalf}\M{\mathcal{W}}\right)\odot\grad c+\grad\cdot\left(\sqrt{2\chi_{0}\vartheta_{0}}\,\M{\mathcal{W}}_{c}\right).
\]
Note that the advective term in the corrector stage can alternatively
be treated using an explicit trapezoidal rule, as in (\ref{eq:impl_trapezoidal}),
without affecting the formal order of accuracy of the scheme.

\begin{algorithm}
\caption{\label{alg:OverdampedConstant}Implicit trapezoidal integrator for
the limiting (overdamped) equation (\ref{eq:limiting_Strato_const}),
as implemented in the IBAMR software framework \cite{IBAMR}.}

\begin{enumerate}
\item In the predictor stage, solve the steady Stokes equation with random
forcing,
\begin{align*}
\grad\pi^{n} & =\eta\grad^{2}\V v^{n}+\grad\cdot\left(\sqrt{\frac{2\eta\, k_{B}T}{\D t\D V}}\,\M W^{n}\right)-\rho\beta c^{n}\V g\\
\grad\cdot\V v^{n} & =0.
\end{align*}

\item Do a predictor step for (\ref{eq:limiting_Strato}) by solving for
$c^{p,n+1}$, 
\[
\frac{c^{p,n+1}-c^{n}}{\D t}=-\V v^{n}\cdot\grad c^{n}+\chi_{0}\grad^{2}\left(\frac{c^{n}+c^{p,n+1}}{2}\right)+\grad\cdot\left(\sqrt{\frac{2\chi_{0}\vartheta_{0}}{\D t\D V}}\,\V W_{c}^{n}\right).
\]

\item Solve the corrector steady Stokes equation
\begin{align*}
\grad\pi^{n+\frac{1}{2}} & =\eta\left(\grad^{2}\V v^{n+\frac{1}{2}}\right)+\grad\cdot\left(\sqrt{\frac{2\eta\, k_{B}T}{\D t\D V}}\,\M W^{n}\right)-\rho\beta\left(\frac{c^{n}+c^{p,n+1}}{2}\right)\V g\\
\grad\cdot\V v^{n+\frac{1}{2}} & =0.
\end{align*}
Note that the same random stress is used here as in the predictor,
so that if there is no gravity, we can set $\V v^{n+\frac{1}{2}}=\V v^{n}$.
In general if an iterative solver is used the solution from the predictor
$\V v^{n}$ should be used as an initial guess to speed up the convergence. 
\item Take a corrector step for concentration to compute $c^{n+1}$,
\[
\frac{c^{n+1}-c^{n}}{\D t}=-\V v^{n+\frac{1}{2}}\cdot\grad\left(\frac{c^{n}+c^{p,n+1}}{2}\right)+\chi_{0}\grad^{2}\left(\frac{c^{n}+c^{n+1}}{2}\right)+\grad\cdot\left(\sqrt{\frac{2\chi_{0}\vartheta_{0}}{\D t\D V}}\,\V W_{c}^{n}\right).
\]
\end{enumerate}
\end{algorithm}

Algorithm \ref{alg:OverdampedConstant} is weakly first-order accurate
for the nonlinear overdamped dynamics, and second-order accurate for
the linearized overdamped dynamics (\ref{eq:linearized_overdamped}).
Therefore, this method ``kills two birds with one stone'' and can
be used for either strong (nonlinear) \cite{DiffusionJSTAT} or weak
(linearized) noise settings. In Section \ref{sec:GRADFLEX} we use
Algorithm \ref{alg:OverdampedConstant} to study the dynamics of the
development of giant concentration fluctuations in the absence of
gravity.

\section{\label{sec:GiantFluct}Giant Concentration Fluctuations}

In this section we apply our temporal integrators to the study of
diffusive mixing in binary liquid mixtures in the presence of a temperature
gradient and gravity. The equations for the velocity $\V v\left(\V r,t\right)$
and mass concentration $c\left(\V r,t\right)$ for a mixture of two
fluids can be approximated as 
\begin{align}
\rho\partial_{t}\V v+\grad\pi= & \eta\grad^{2}\V v+\grad\cdot\left(\sqrt{2\eta k_{B}T_{0}}\,\M{\mathcal{W}}\right)-\rho\beta c\V g\label{eq:LLNS_comp_v_simp}\\
\grad\cdot\V v= & 0\nonumber \\
\partial_{t}c+\V v\cdot\grad c= & \chi\grad\cdot\left(\grad c+c\left(1-c\right)S_{T}\grad T\right),\label{eq:LLNS_comp_c_simp}
\end{align}
where $\M{\mathcal{W}}$ denotes white-noise stochastic forcing for
the thermal fluctuations and $\V g$ is gravity. Here we have ignored
the stochastic forcing term $\grad\cdot\left(\sqrt{2\chi_{0}\vartheta_{0}}\,\M{\mathcal{W}}_{c}\right)$,
which is responsible for equilibrium fluctuations in the concentration,
since our focus will be on the much larger nonequilibrium fluctuations
induced by the coupling to the velocity equation via the advective
term $\V v\cdot\grad c$. The shear viscosity $\eta=\nu\rho$, mass
diffusion coefficient $\chi$, solutal expansion coefficient $\beta$,
and Soret coefficient $S_{T}$, are assumed to be given material constants
independent of concentration. Furthermore, the density $\rho$ is
taken to be constant in a Boussinesq approximation. Temperature fluctuations
are not considered in a large Lewis number (very fast temperature
dynamics) approximation \cite{GiantFluctFiniteEffects}. We assume
that the applied temperature gradient $\grad T$ is weak and approximate
$T\approx T_{0}=\mbox{const}$. In principle there is no difficulty
in making the temperature be spatially-dependent, however, our simplifying
approximation is justified because the typical relative temperature
difference across the sample is not large.

These equations are extremely difficult to integrate numerically for
large Schmidt number, $S_{c}=\nu/\chi\gg1$, if one wants to get dynamics
correctly. Therefore, we actually take a limit of the above equations
$S_{c}\rightarrow\infty$ and integrate the resulting dynamics numerically.
In a linearized setting, $c=\bar{c}+\d c$ and $\V v\equiv\d{\V v}$,
the overdamped equations are written in \cite{GiantFluctFiniteEffects}
as a large Schmidt and large Lewis number approximation to the complete
system of equations, 
\begin{align}
\grad\pi= & \eta\grad^{2}\V v+\grad\cdot\left(\sqrt{2\eta k_{B}T_{0}}\,\M{\mathcal{W}}\right)-\rho\beta\left(\d c\right)\V g\label{eq:vc_linearized_overdamped}\\
\grad\cdot\V v= & 0\nonumber \\
\partial_{t}\left(\d c\right)+\V v\cdot\V h= & \chi\grad^{2}\left(\d c\right),\nonumber 
\end{align}
where $\V h=\grad\bar{c}$ is the concentration gradient imposed by
the applied temperature gradient. Theoretical analysis is based on
the simplified linearized equations (\ref{eq:vc_linearized_overdamped})
under the assumption that the appplied gradient $\V h$ is constant
and weak. The theory predicts the \emph{steady-state} spectrum of
the concentration fluctuations at wave number $\V k$ to be \cite{GiantFluctFiniteEffects}

\begin{equation}
S\left(\V k\right)=\av{\left(\widehat{\delta c}\left(\V k\right)\right)\left(\widehat{\delta c}\left(\V k\right)\right)^{\star}}=\frac{k_{B}T_{0}}{\left(\eta\chi k_{\perp}^{4}-g\rho\beta h_{\parallel}\right)}\, h_{\parallel}^{2},\label{eq:S_c_c}
\end{equation}
where \textbf{$\perp$} and $\parallel$ denote the perpendicular
and parallel component relative to gravity, respectively. The characteristic
$k_{\perp}^{4}$ power-law divergence of the spectrum at large wavenumbers
is a signature of long-ranged nonequilibrium fluctuations and leads
to a dramatic increase in the magnitude and correlation length of
the fluctuations compared to systems in thermodynamic equilibrum;
this effect has been termed \emph{giant fluctuations} \cite{GiantFluctuations_Nature,FractalDiffusion_Microgravity}.
The expression (\ref{eq:S_c_c}) shows that fluctuations at wavenumbers
below the critical (rollover) wave number $k_{c}^{4}=g\rho\beta h_{\parallel}/\left(\eta\chi\right)$
are suppressed by gravity. Henceforth we will assume that the gradient
is parallel to gravity, $h_{\parallel}=h$.

For the dynamics of vertically-averaged concentration, i.e., for $k_{\parallel}=0$,
$k_{\perp}=k$, the linearization (\ref{eq:vc_linearized_overdamped})
predicts an exponential time correlation function
\begin{equation}
S\left(k,t\right)=\av{\left(\widehat{\delta c}\left(k,t\right)\right)\left(\widehat{\delta c}\left(k,0\right)\right)^{\star}}=S\left(k\right)e^{-t/\tau}\label{eq:S_k_t_overdamped}
\end{equation}
with decay time
\begin{equation}
\tau^{-1}=\chi k^{2}\left[1+\left(\frac{k_{c}}{k}\right)^{4}\right].\label{eq:tau_overdamped}
\end{equation}
The relaxation has a minimum at $k=k_{c}$ with value $\tau_{\text{min}}^{-1}=2\chi k_{c}^{2}$.
For smaller wavenumbers $\tau$ becomes the smallest time scale and
limits the stability of the simulations. As we discuss in Section
\ref{sec:SORET}, at small wavenumbers the separation of time scales
used to justify the overdamped limit fails and the fluid inertia has
to be taken into account. In the absence of gravity, however, as we
discuss in Section \ref{sec:GRADFLEX}, the separation of time scales
is uniform across all length scales and the overdamped limit can be
used.

\subsection{\label{sec:GRADFLEX}Giant Fluctuations in Microgravity}

In this section we perform computer simulations of diffusive mixing
in microgravity, recently studied aboard a satellite in orbit around
the Earth during the GRADFLEX experiment \cite{FractalDiffusion_Microgravity}.
The experimental configuration consists of a dilute solution of polystyrene
in toluene with average concentration $\vartheta_{0}=0.018$, confined
between two parallel transparent plates that are a distance $H=1\mbox{mm}$
apart. A temperature gradient $\grad T=\D T/H$ is imposed along the
$y$ axes via the plates at time $t=0$. At long times, the weak temperature
gradient leads to a strong concentration gradient $\grad\bar{c}=\bar{c}S_{T}\grad T$
due to the Soret effect, giving rise to an exponential steady-state
concentration profile $\bar{c}(y)$. Quantitative shadowgraphy is
used to observe and measure the strength of the fluctuations in the
concentration around $\bar{c}$ via the change in the refraction index.
Some of us already modeled the \emph{steady-state} fluctuations in
the GRADFLEX experiment using the inertial equations (\ref{eq:LLNS_comp_v_simp},\ref{eq:LLNS_comp_c_simp})
in previous work \cite{LLNS_Staggered}. Here we extend this study
to also model the \emph{dynamics} of the concentration fluctuations
following the time when the temperature gradient is first applied
(to a uniform sample), before the steady state is reached. In future
work we will compare our numerical results to experimental measurements.

In the actual experiments reported in Ref. \cite{FractalDiffusion_Microgravity},
concentration diffusion is much slower than momentum diffusion, corresponding
to Schmidt number $S_{c}=\nu/\chi\approx3\cdot10^{3}$. This level
of stiffness makes direct simulation of the temporal dynamics of the
fluctuations infeasible, as long averaging is needed to obtain accurate
steady-state spectra, especially for small wavenumbers. In order to
bypass this problem, in our previous work \cite{LLNS_Staggered} we
artificially increased $\chi$ and decreased $\nu$ to reduce the
Schmidt number, while keeping the product $\chi\nu$ fixed. This is
exactly the scaling in which one can formally derive the limiting
overdamped dynamics (\ref{eq:limiting_Strato}) \cite{DiffusionJSTAT},
and from (\ref{eq:S_c_c}) we see that the static structure factor
depends only on the product $\chi\nu$ when $\nu\gg\chi$. In fact,
artificially decreasing the Schmidt number while keeping $\chi\nu$
fixed, and rescaling time appropriately, can be seen as an instance
of the \emph{seamless} multiscale method presented in Ref. \cite{SeamlessMultiscale}.
Here we take the overdamped limit and solve the limiting equations
numerically, thus completely avoiding the stiffness issue. This allows
us to take a much larger time step size, related to the time scale
of mass diffusion, rather than the fast momentum diffusion.

For the GRADLEX experiments we can assume $c\ll1$ and thus $c(1-c)\approx c$
to make the Soret term linear in concentration and treat the Soret
flux as advection by a Soret velocity $\V v_{s}=-\chi S_{T}\grad T$,
to obtain the concentration equation employed in our numerical method,
\begin{equation}
\partial_{t}c+\grad\cdot\left(c\left(\V v-\chi S_{T}\grad T\right)\right)=\chi\grad^{2}c.\label{eq:c_GRADFLEX}
\end{equation}
We integrate the overdamped limit of (\ref{eq:LLNS_comp_v_simp},\ref{eq:c_GRADFLEX})
for $g=0$ in time using Algorithm \ref{alg:OverdampedConstant}.
The spatial discretization and the physical parameters are essentially
identical to those used in incompressible simulations in our previous
work, see Section V in \cite{LLNS_Staggered}. An important improvement
is that we now handle both the Soret term and the boundary condition
for concentration implicitly, thus ensuring strict conservation of
the total solutal mass. The domain is periodic in the directions parallel
to the boundaries. At the top and bottom boundaries a no-flux boundary
condition is imposed for the concentration, and a no-slip boundary
condition is imposed for velocity. The observed light intensity, once
corrected for the optical transfer function of the equipment, is proportional
to the intensity of the fluctuations in the concentration averaged
along the gradient, 
\[
c_{\perp}(x,z;t)=H^{-1}\int_{0}^{H}c(x,y,z;t)dy,
\]
and this is the main quantity of interest in our simulations. What
is actually measured experimentally is the static structure factor,
which is the Fourier transform $\widehat{\d c}_{\perp}$ of the concentration
fluctuations averaged along the gradient direction, 
\[
S\left(k_{x},k_{z};t\right)=\av{\left(\widehat{\d c}_{\perp}\right)\left(\widehat{\d c}_{\perp}\right)^{\star}}.
\]

Because of the increase in the time step afforded by the use of the
overdamped integrator, we are able here to perform fully three-dimensional
simulations on a domain of dimensions $\left(4\times1\times4\right)\,\text{mm}$,
discretized on a $256\times64\times256$ grid with uniform grid spacing
$\D x=1/64\,\text{mm}$. Here the thickness of the sample is $H=1\,\text{mm}$
and corresponds to the experimental setup, and the lateral extend
is set to $L=4\,\text{mm}$. The structure factors $S\left(k_{x},k_{z};t\right)$
were averaged radially to obtain $S\left(k;t\right)$, where $k=\sqrt{k_{x}^{2}+k_{z}^{2}}$.
Note that in this case it is possible to obtain the same results using
two-dimensional simulations ($k_{z}=0$) because of the symmetries
of the linearized equations, nevertheless, we chose to obtain three-dimensional
results directly comparable to experiments. The key physical parameters
are $\rho=0.858\,\mbox{g}/\mbox{cm}^{3}$, $\chi=1.97\cdot10^{-6}\,\mbox{cm}^{2}/\mbox{s}$,
$\nu=6.07\cdot10^{-3}\,\mbox{cm}^{2}/\mbox{s}$, $T_{0}=300\,\mbox{K}$,
$S_{T}=0.06486\,\mbox{K}^{-1}$, and the temperature difference is
$\D T=17.4\,\mbox{K}$. Additional details of the experimental setup
and parameters are given in Ref. \cite{FractalDiffusion_Microgravity}.
The time step size was $\D t\approx10\,\mbox{s}$, corresponding to
a diffusive Courant number $\chi\D t/\D x^{2}\approx8$, and the results
were averaged over 32 simulations. To obtain the static structure
factor at steady state, we used a time step size that was twice larger,
and averaged over 32 runs of length $140,000\,\mbox{s},$ skipping
the initial $14,000\,\mbox{s}$ in the analysis.

\begin{figure*}
\begin{centering}
\includegraphics[width=0.49\textwidth]{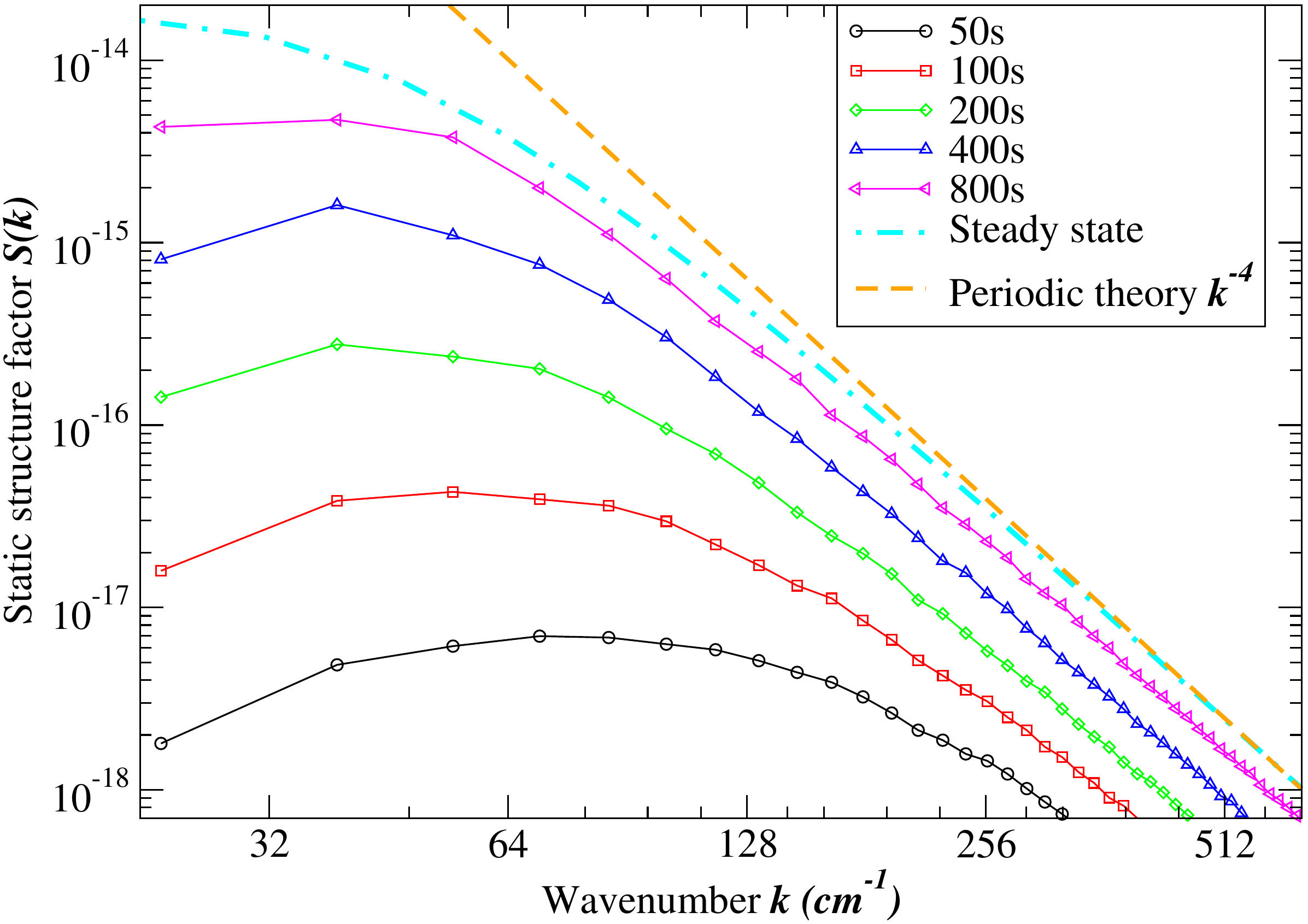}\includegraphics[width=0.49\textwidth]{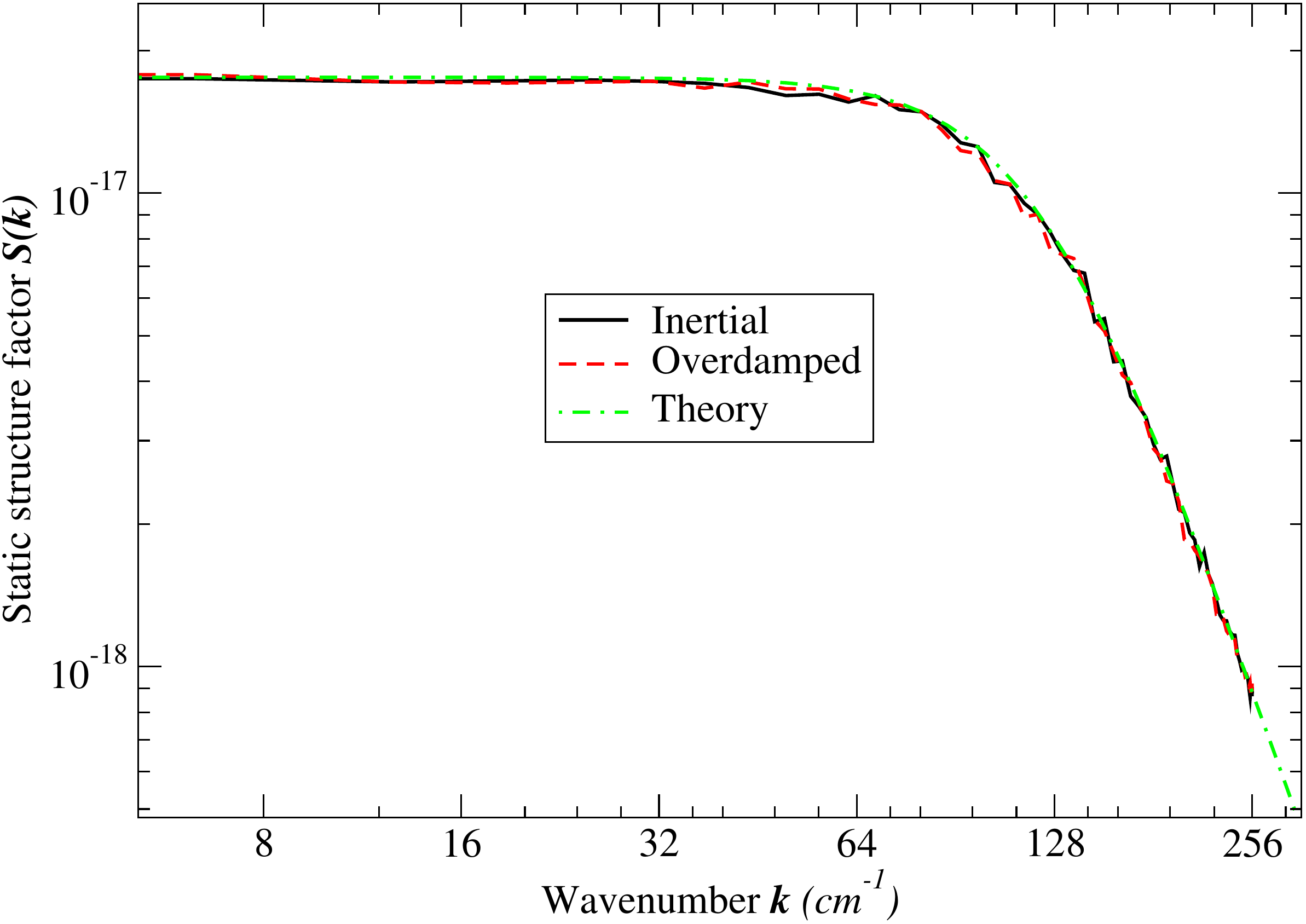}
\par\end{centering}

\caption{\label{fig:S_k}(Color online) (Left) Time evolution of the static
structure factor $S(k;t)$ in the GRADFLEX experiment \cite{FractalDiffusion_Microgravity}.
The steady-state spectrum, studied in more detail in Section V in
\cite{LLNS_Staggered}, is also shown, and compared to the simple
quasi-periodic theory (\ref{eq:S_c_c}). At smaller wavenumbers the
steady-state fluctuations are damped due to confinement effects. (Right)
Static structure factor $S(k)$ of concentration fluctuations in a
binary liquid mixture subjected to a steady temperature gradient \cite{SoretDiffusion_Croccolo}.
In order to account for spatial discretization artifacts, on the $x$
axes we show the modified wave number $\sin\left(k\D x/2\right)/\left(\D x/2\right)$
\cite{LLNS_Staggered}. The theory (\ref{eq:S_c_c}) is shown for
comparison.}
\end{figure*}

In the left panel of Fig. \ref{fig:S_k} we show the static structure
factor as a function of time, along with the steady state asymptotic
value. We observe that it takes a long time, on the time scale of
the macroscopic diffusive mixing, to actually establish the steady
state. This is in contrast to experiments performed in gravity, where
gravity accelerates the dynamics of the concentration fluctuations
at the smaller wavenumbers, as seen in (\ref{eq:tau_overdamped}).
This was used in Refs. \cite{GiantFluctuations_Theory,GiantFluctuations_Cannell}
to construct a simple but approximate theory for the time evolution
of the static structure factor during free diffusive mixing of two
miscible liquids. Such a simple theory cannot be developed in microgravity
because there is no separation of time scales between the dynamics
of the mean and the dynamics of the fluctuations around the mean,
and computer simulations become indispensable.

\subsection{\label{sec:SORET}Giant Fluctuations in Earth Gravity}

In this section we consider giant fluctuations in the presence of
gravity. We model the experiments used in Ref. \cite{SoretDiffusion_Croccolo}
to measure the Soret and diffusion coefficients in binary mixtures
using a setup similar to that in the GRADFLEX experiment described
in the previous section. A notable difference is that the average
concentration $\vartheta_{0}\approx0.5$ is much larger than in the
GRADFLEX setup, and only a small relative concentration difference
is induced by an applied modest temperature gradient. We therefore
approximate $c\left(1-c\right)\approx\vartheta_{0}\left(1-\vartheta_{0}\right)$
in (\ref{eq:LLNS_comp_c_simp}), giving a constant mass Soret flux.
In this example, because we want to accurately resolve the decay of
the time correlation functions over several orders of magnitude, we
perform two dimensional simulations. As we already explained, there
is no difference between two and three dimensional simulations because
of the simple one-dimensional geometry of the deterministic solution
\cite{FluctHydroNonEq_Book}.

By linearizing the inertial dynamics (\ref{eq:LLNS_comp_v_simp},\ref{eq:LLNS_comp_c_simp})
and taking a spatio-temporal Fourier transform we can obtain an approximate
theory for the spatio-temporal correlation functions for the concentration
fluctuations. This straightforward analysis predicts that the static
factor $S\left(k\right)$ continues to follow (\ref{eq:S_c_c}) if
$\nu\gg\chi$, but the time correlation function, unlike the overdamped
result (\ref{eq:S_k_t_inertial}), is found to be a sum of two modes
\begin{equation}
S(k,t)=S_{0}\exp\left(-\frac{t}{\tau_{1}}\right)+\left(S\left(k\right)-S_{0}\right)\exp\left(-\frac{t}{\tau_{2}}\right),\label{eq:S_k_t_inertial}
\end{equation}
where we omit the long formula for $S_{0}$ and just quote the relaxation
times
\[
\tau_{1/2}^{-1}=\frac{1}{2}\left(\nu+\chi\right)k^{2}\pm\frac{1}{2}\,\sqrt{k^{4}\left(\nu-\chi\right)^{2}-4\beta gh}.
\]
In the limit $\nu/\chi\rightarrow\infty$, we get the diffusive relaxation
time (\ref{eq:tau_overdamped}) if we use the minus sign, $\tau_{\chi}^{-1}\approx\chi k^{2}$,
and for the plus sign we get the viscous relaxation time
\begin{equation}
\tau_{\nu}^{-1}=\nu k^{2}\left[1-\frac{\beta gh}{\nu\chi k^{4}}\right]\approx\nu k^{2}.\label{eq:tau_overdamped-1}
\end{equation}
where $\nu=\eta/\rho$.

Note however that the relaxation times become complex-valued for
\[
k_{p}\lesssim\left(\frac{4\beta gh}{\nu^{2}}\right)^{\frac{1}{4}}=\left(4\frac{\chi}{\nu}\right)^{\frac{1}{4}}k_{c}=\sqrt{2}\, S_{c}^{-\frac{1}{4}}k_{c},
\]
indicating the appearance of \emph{propagative} rather than diffusive
modes for small wavenumbers. Because of the fourth root power, for
realistic values of $S_{c}\sim10^{4}$, propagative modes appear at
wavenumbers that are, in principle, observable in experiments via
very low-angle light scattering and shadowgraph techniques, although,
to our knowledge, experimental observation of propagative modes has
only been reported for temperature fluctuations \cite{TemperatureGradient_Cannell}.
This shows that there is a \emph{qualitative} difference between the
inertial and overdamped dynamics in this example. This comes because
of the lack of separation between the relaxation times associated
with mass ($\tau_{\chi}$) and momentum diffusion ($\tau_{\nu}$)
for wavenumbers $k\lesssim k_{p}$. In order to obtain accurate results
over the whole range of wavenumbers observed in experiments, we need
to account for the fluid inertia and integrate the system in time
using Algorithm (\ref{alg:InertialConstant}). For comparison we also
numerically take the overdamped limit and use Algorithm (\ref{alg:OverdampedConstant})
for the temporal integration.

In our simulations we use a grid of $128\times128$ cells, with grid
spacing $\D x=\D y=1/128$cm. This corresponds to thickness of the
sample of $H=L=1\mbox{cm}$. The physical parameters correspond to
the THN-C12 mixture studied in \cite{SoretDiffusion_Croccolo}, $\rho=0.8407\,\mbox{g}/\mbox{cm}^{3}$,
$\chi=6.21\cdot10^{-6}\,\mbox{cm}^{2}/\mbox{s}$, $\nu=1.78\cdot10^{-2}\,\mbox{cm}^{2}/\mbox{s}$,
$T_{0}=300\,\mbox{K}$, $S_{T}=9.5\cdot10^{-3}\,\mbox{K}^{-1}$, $\beta=0.27$,
$g=981\,\mbox{cm}/\mbox{s}^{2},$ and the temperature difference across
the sample is $\D T=40\,\mbox{K}$. To obtain the time-correlation
functions we analyze a single run corresponding to $3,125\,\mbox{s}$
of physical time, skipping the initial $625\,\mbox{s}$. The time
step used in both the inertial and overdamped integrators is $\D t=5\cdot10^{-3}$s,
giving a viscous Courant number $\nu\D t/\D x^{2}=1.5$. This means
that the viscous dynamics is well-resolved by this small time step
and there is no real benefit from using the overdamped equations.
A larger time step cannot be used here even in the overdamped limit
because the relaxation time (\ref{eq:tau_overdamped}) for the smallest
wave number $k\approx6.28\,\text{cm}^{-1}$ is $\tau=0.025\,\text{s}$,
and therefore resolving the dynamics of the concentration requires
a rather small time step. 

\begin{figure*}
\begin{centering}
\includegraphics[width=0.85\textwidth]{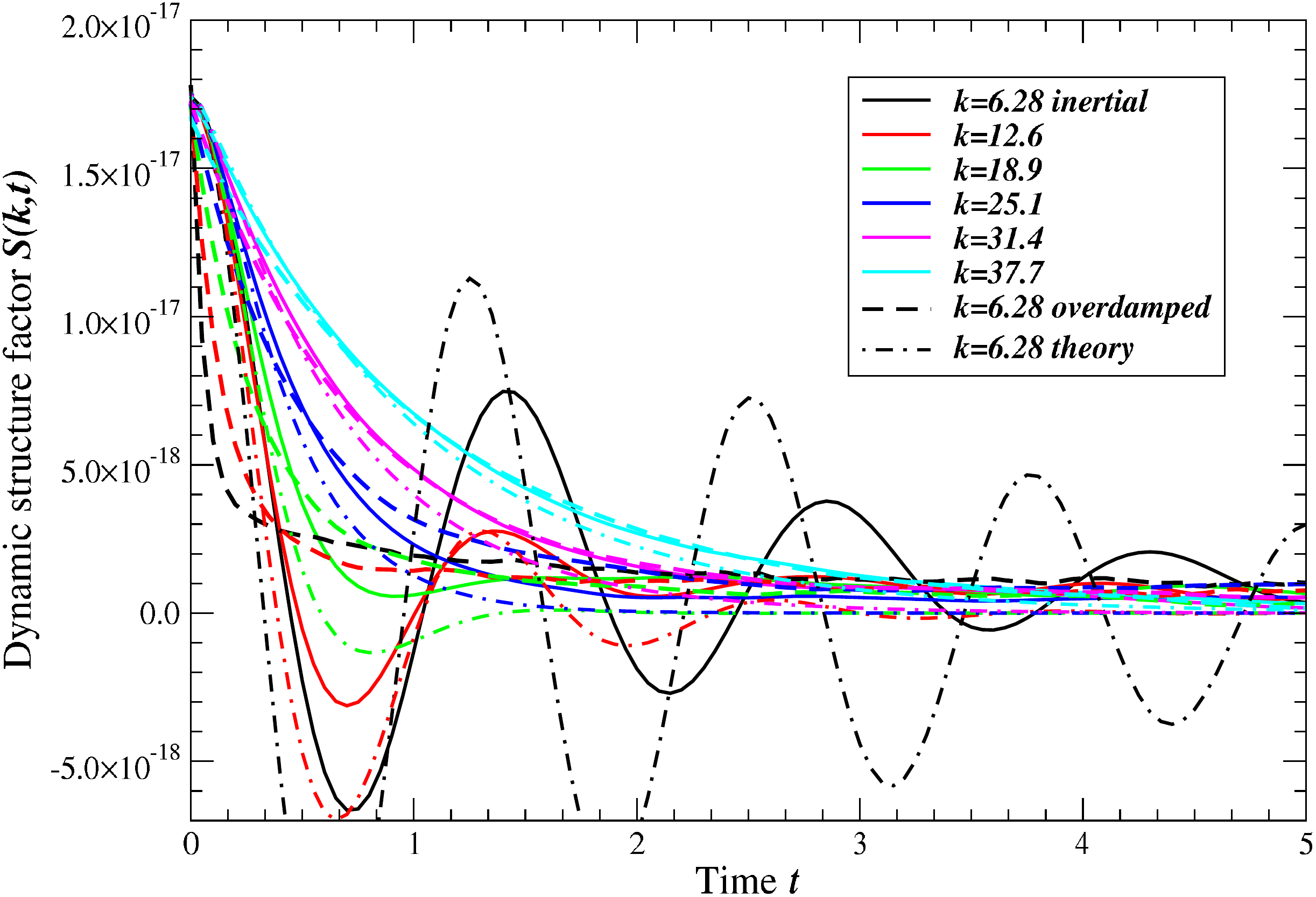}
\par\end{centering}

\caption{\label{fig:SORET}(Color online) Time correlation functions $S(k,t)$
of concentration fluctuations for concentration fluctuations in a
binary liquid mixture subjected to a temperature gradient \cite{SoretDiffusion_Croccolo},
for the first few wavenumbers (listed in the legend in $\mbox{cm}^{-1}$).
Solid lines are obtained using the inertial equations, while dashed
lines of the same color are obtained using the overdamped approximation.
For the smallest two wavenumbers clear oscillations are observed indicating
the appearance of propagative modes. These are not captured correctly
in the overdamped approximation, and are only qualitatively described
by the simple theory (\ref{eq:S_k_t_inertial}), shown with dashed-dotted
lines.}
\end{figure*}

In the right panel of Fig. \ref{fig:S_k} we show numerical results
for the static structure factor $S(\V k)$, while in Fig. \ref{fig:SORET}
we show results for the dynamic structure factor $S(\V k,t)$. While
the static factor shows no difference between the inertial and overdamped
integrators, the dynamic factor clearly shows the appearance of oscillations
(propagative modes) for the smallest wavenumbers when inertia is accounted
for. For comparison, in Fig. \ref{fig:SORET} we also plot the theory
(\ref{eq:S_k_t_inertial}). We see a qualitative but not a quantitative
agreement between the numerical results and the theory. This mismatch
can be attributed to the effects of confinement by the two no-sip
boundaries, which is not taken into account in the simple quasi-periodic
theory used to derive (\ref{eq:S_k_t_inertial}). This effect becomes
stronger the smaller the value of the dimensionless product $kH$
is. Constructing analytical theories in the presence of confinement
and inertial effects is quite challenging and computer simulations
are required to study these effects, which we will explore in more
detail in future publications.

\section{\label{sec:Conclusions}Conclusions}

Continuing on our previous work \cite{DFDB}, we constructed a general
class of mixed explicit-implicit predictor-corrector schemes for integrating
Langevin equations, and recommended two specific schemes that we previously
proposed in \cite{DFDB} in the context of additive-noise Langevin
equations. The first is a fully explicit midpoint rule, and the second
is a semi-implicit scheme in which some of the terms are treated using
an implicit trapezoidal rule and the rest are treated using an explicit
trapezoidal rule. Here we showed how to add stochastic forcing terms
to these schemes that ensure the following key properties: the schemes
are second-order weakly accurate for linearized Langevin equations,
and they are weakly first-order accurate for multiplicative kinetic
noise. In particular, we discussed how to obtain all of the required
stochastic drift terms without evaluating derivatives, using random
finite differences that give the required stochastic drift in expectation.

The key idea in our approach was to discretize the original nonlinear
Langevin equations with sufficiently weak noise, and let the algorithm
do the linearization for us, without us having to even write the linearized
equations explicitly. This has numerous advantages over the alternative
approach of splitting the variables into a deterministic mean plus
fluctuations and then linearizing the equations manually. Firstly,
the linearization needs to be done around an unknown time-dependent
solution of the deterministic equations, which must itself be computed
numerically in general. Secondly, the linearized equations typically
have many more terms than the nonlinear equations, and do not obey
conservation laws, leading to violation of conservation for the overall
solution (mean+fluctuations). Lastly, our algorithms can, with care,
also be used to integrate genuinely nonlinear Langevin equations,
and thus numerically access the importance of terms neglected by linearization
\cite{DiffusionRenormalization,DiffusionJSTAT}. Integrating the linearized
equations to second-order weak accuracy proved to be relatively simple.
Essentially, all that is required is to evaluate the noise amplitude
in the corrector step at the correct value for the time-dependent
deterministic solution, e.g., at the midpoint or end-point of the
time step. In the general nonlinear setting, by contrast, obtaining
second-order accuracy with derivative-free schemes is rather nontrivial
\cite{WeakSecondOrder_RK}.

We also proposed predictor-corrector schemes for integrating systems
of Langevin equations containing a fast and a slow variable, in the
limit of infinite separation of time scales. We limited our attention
here to the simple but common case of the fast variable entering only
linearly; the general case is much more subtle \cite{AdiabaticElimination_1}.
Our predictor-corrector schemes discretize the original equations
but without the time derivative term in the equation for the fast
variable, giving an effective integrator for the overdamped equations
without ever explicitly even writing the limiting dynamics. One way
to think about our integrators is as applying a backward Euler method
to the fast variable, since this method finds a steady state solution
of the fast variable in the limit when the time step size is much
larger than the internal time scale of the fast variable. An essential
difficulty in the construction of integrators for the nonlinear overdamped
equations is capturing the stochastic drift terms that arise due to
the kinetic interpretation of the noise. We accomplished this goal
by using a combination of implicit and explicit random finite differences.

The integrators proposed here handle fast-slow systems by taking the
limit of infinite separation of variables to eliminate the fast variable.
Such an asymptotic limit is often a good approximation, as we illustrated
in the past when modeling diffusion in liquids \cite{DiffusionJSTAT},
and also here by modeling the development of giant fluctuations in
microgravity during the GRADFLEX experiment \cite{FractalDiffusion_Microgravity}.
However, we also studied here the dynamics of giant fluctuations in
Earth gravity and found that the assumption of \emph{uniform} separation
of time scales between the fast velocity and slow concentration fails
in practice for sufficiently small wavenumbers. In this paper, we
resorted to integrating the original (inertial) equations using a
sufficiently small time step. It remains an important challenge for
the future to develop \emph{uniformly accurate} temporal integrators
that remain weakly first-order accurate even when there is incomplete
separation of time scales, while still allowing the use of large time
step sizes, on the time scale of the slow variable. Such integrators
must likely rely on the linearity of the equation and exponential
integrators \cite{SDEWeakExponential,StochasticExponential,StochasticImmersedBoundary}
for the fast variable. The key difficulty, which we hope to address
in future work, is to correctly capture the effects of the fast velocity
on the slow concentration that arise due to the nonlinear advective
term $\V v\cdot\grad c$ \cite{DiffusionJSTAT}.
\begin{acknowledgments}
We would like to thank John Bell, Alejandro Garcia, Andy Nonaka, and
Jonathan Goodman for numerous enlightening discussions. We are indebted
to Fabrizio Croccolo and Alberto Vailati for their assistance in setting
up the simulations of giant fluctuations, and helpful comments on
this manuscript. This material is based upon work supported by the
U.S. Department of Energy Office of Science, Office of Advanced Scientific
Computing Research, Applied Mathematics program under Award Number
DE-SC0008271. A. Donev and Y. Sun were supported in part by the National
Science Foundation (NSF) under grant DMS-1115341. B. Griffith and
S. Delong acknowledge research support from the NSF under award OCI
1047734. E. Vanden-Eijnden and S. Delong were supported by the DOE
office of Advanced Scientific Computing Research under grant DE-FG02-88ER25053.
Additional support for E. Vanden-Eijnden was provided by the NSF under
grant DMS07-08140 and by the Office of Naval Research under grant
N00014-11-1-0345.
\end{acknowledgments}
\appendix

\section*{Appendix}

\section{Accuracy for time dependent noise\label{sec:Appendix-time-dependent}}

In this appendix, we show that the scheme (\ref{eq:predcorr_scheme})
successfully captures the time dependence of $\M K$ in the equation
(\ref{eq:TimeDepNoise}) to second order. The general theory of weak
accuracy for stochastic integrators is well-established and reviewed,
for example, in Section 2.2 of \cite{MilsteinSDEBook}. The key result
is that, under certain assumptions, second-order weak accuracy is
achieved if the first $5$ moments of the numerical increment $\D{\V x}^{n}=\V x^{n+1}-\V x^{n}$
match the moments of the true increment $\V x\left(n\D t+\D t\right)-\V x\left(n\D t\right)$
with error no greater than $O\left(\D t^{3}\right)$. The time dependence
of the noise amplitude only adds one additional term (beyond constant
noise) that a second-order integrator must account for, which we find
from the expansion 
\begin{align}
\int_{t_{n}}^{t_{n+1}}\M K(s)\, d\V{\mathcal{B}}(s)= & \,\M K(t_{n})\int_{t_{n}}^{t_{n+1}}\, d\V{\mathcal{B}}(s)+\M{\partial K}(t_{n})\int_{t_{n}}^{t_{n+1}}(s-t_{n})\, d\V{\mathcal{B}}(s)\label{eq:noise_time_expansion-1}\\
 & +\M{\partial^{2}K}(t_{n})\int_{t_{n}}^{t_{n+1}}\frac{(s-t_{n})^{2}}{2}\, d\V{\mathcal{B}}(s)+O\left(\Delta t^{7/2}\right).\nonumber 
\end{align}
Here $\M{\partial K}$ indicates a time derivative of $\M K$, and
likewise $\M{\partial^{2}K}$ indicates a second derivative in time.
The first term in the second line is order $\D t^{5/2}$ and mean
zero, so it will not affect second order weak accuracy. Likewise,
terms which arise from the time dependence of $\M K$ in the Taylor
expansion of the deterministic part of (\ref{eq:TimeDepNoise}) are
all order greater than two and a half and mean zero. Therefore, the
only additional expression which must be accounted for in the temporal
integrator is the second term on the first line of equation (\ref{eq:noise_time_expansion-1}).
When matching increments for second order weak accuracy, this term
appears only in the second moment of the increment, and only in the
cross term with $\M K(t_{n})\int_{t_{n}}^{t_{n+1}}d\V{\mathcal{B}}(s)$,
\begin{align}
E\left[\left(\V x(t_{n+1})-\V x(t_{n})\right)\left(\V x(t_{n+1})-\V x(t_{n})\right)^{\star}\right]= & \ldots+\left((\M{\partial K})\M K^{\star}+\M K(\M{\partial K}){}^{\star}\right)\frac{\D t^{2}}{2}+O(\D t^{3})\label{eq:second_moment-1}
\end{align}
where the terms that are present in the constant-noise case already
appear in (16) in \cite{DFDB} and are denoted with ellipses. 

In order to maintain second order weak accuracy, the additional term
involving $\M{\partial K}$ in (\ref{eq:second_moment-1}) is matched
in the temporal integrator by the way in which the noise is evaluated
in the corrector step of (\ref{eq:predcorr_scheme}), 
\begin{align}
\V x^{n+1}-\V x^{n}= & \ldots+\left((1-w_{7})\M K^{n}+w_{7}\M K^{(p)}\right)\left(\sqrt{w_{2}\D t}\,\V W^{n,1}+\sqrt{(1-w_{2})\D t}\,\M W^{n,2}\right),\label{eq:corrector_noise}
\end{align}
where deterministic terms are denoted by elipses. Taylor expanding
$\M K^{(p)}$ in (\ref{eq:corrector_noise}) gives the following two
terms,
\begin{align}
\V x^{n+1}-\V x^{n}= & \ldots+\D t^{1/2}\,\M K^{n}\left(\sqrt{w_{2}}\,\V W^{n,1}+\sqrt{1-w_{2}}\,\V W^{n,2}\right)\nonumber \\
 & +\D t^{3/2}w_{2}w_{7}\M{\partial K}\left(\sqrt{w_{2}}\,\V W^{n,1}+\sqrt{1-w_{2}}\,\V W^{n,2}\right)+O(\D t^{5/2}).\label{eq:noise_expansion}
\end{align}
The terms on the right hand side create the correct cross term in
the second moment of the increment as long as $w_{2}w_{7}=1/2$. Therefore,
scheme (\ref{eq:predcorr_scheme}) is second order weakly accurate
even for time-dependent additive noise.

\section{\label{Appendix-Accuracy}Second-order weak integrators for linearized
fluctuating hydrodynamics}

Here we show that the scheme (\ref{eq:general_linearizer}) is second
order accurate for the linearized equations (\ref{eq:dxbar_dt},\ref{eq:dx_dt_lin}).
For the scheme to be weakly second order accurate, we need the difference
between the first five moments of the discrete and continuous increments
for the composite variable $\left(\bar{\V x}(t),\,\d{\V x}(t)\right)$
to be $O\left(\D t^{3}\right)$ \cite{MilsteinSDEBook}. The scheme
(\ref{eq:general_x_bar}) for the mean is deterministic and standard
Taylor series analysis shows that it is second-order accurate. Furthermore,
because $\V{\bar{x}}$ is deterministic, there are no cross-correlations
between increments of $\bar{\V x}$ and $\d{\V x}$, and therefore
moments that involve both variables are guaranteed to match to the
same order as the increments in $\d{\V x}$.

In order to examine the moments of the increment in $\d{\V x}$, let
us re-write equation (\ref{eq:dx_dt_lin}) in differential and index
notation, 
\begin{align}
d(\delta x_{i})= & M_{ij}\left(\bar{x}\left(t\right)\right)\delta x_{j}\, dt+K_{ij}\left(\bar{x}\left(t\right)\right)d\mathcal{B}_{j}\left(t\right),\label{eq:linearization_index}
\end{align}
where as before $M_{ij}=H_{ij}+\partial_{j}(H_{ik})\bar{x}_{k}+\partial_{j}(h_{i})$,
giving
\[
\partial_{k}\left(M_{ij}\right)=\partial_{k}\left(H_{ij}\right)+\partial_{kj}\left(H_{il}\right)\bar{x}_{l}+\partial_{j}\left(H_{ik}\right)+\partial_{jk}\left(h_{i}\right).
\]
Using the deterministic equation
\[
d\bar{x}_{k}/dt=H_{kl}\bar{x}_{l}+h_{k},
\]
it can be shown that over a time interval $\Delta t=t^{n+1}-t^{n}$,
the random variable $\delta x_{i}$ has the following continuous increment,
\begin{align}
\Delta\left(\delta x_{i}\right)=\d x_{i}(t^{\prime})-\d x_{i}\left(t\right)= & \D t\, M_{ij}\d x_{j}+K_{ij}\int_{t^{n}}^{t^{n+1}}d\mathcal{B}_{j}(s)\label{eq:fluctuation_continuous_increment}\\
 & +\frac{\D t^{2}}{2}\partial_{k}\left(M_{ij}\right)\d x_{j}\left(\frac{d\bar{x}_{k}}{dt}\right)+\frac{\D t^{2}}{2}M_{ij}M_{jk}\d x_{k}\nonumber \\
 & +\partial_{k}\left(K_{ij}\right)\left(\frac{d\bar{x}_{k}}{dt}\right)\int_{t^{n}}^{t^{n+1}}(s-t)\, d\mathcal{B}_{j}(s)+M_{ij}K_{jk}\int_{t^{n}}^{t^{n+1}}\int_{t}^{s}\, d\mathcal{B}_{k}(r)\, ds+O(\D t^{5/2})\nonumber 
\end{align}
where in this equation and in the following, $d\bar{x}_{k}/dt$, as
well as $\M M,\,\M H,\,\M K$, and $\V h$ and their derivatives,
are all evaluated at the beginning of the time step.

In order to compare this to the discrete increment, we need to perform
a Taylor expansion of every term in equation (\ref{eq:general_dx_np1})
that is not already evaluated at $\bar{\V x}^{n}$. These expansions
never need to include terms of higher order than $\D t^{3/2}$, considering
that all terms in (\ref{eq:general_dx_np1}) are already at least
order $\D t^{1/2}$ (any term that is $O(\D t^{5/2})$ is mean zero
and thus irrelevant). Additionally, the only term to be expanded which
is order $\D t^{1/2}$ before expansion is the stochastic term $w_{7}K_{ij}^{p}\left(\sqrt{w_{2}\D t}\, W_{j}^{1,n}+\sqrt{(1-w_{2})\D t}\, W_{j}^{2,n}\right)$,
which depends only on the deterministic variable $\bar{\V x}$. Thus
an expansion of $K_{ij}^{p}$ to order $\D t^{3/2}$ only includes
first order terms. All other terms in (\ref{eq:general_dx_np1}) to
be expanded are of order $\D t$ or greater before expansion, so in
fact we only need to consider first order expansions for all quantities
evaluated at $\bar{\V x}^{p}$ or $\bar{\V x}^{n+1}$ in the discrete
increment.

For $\delta\V x,$ we expand the update (\ref{eq:general_dx_p},\ref{eq:general_dx_np1})
to first order, substitute the result in the corrector update (\ref{eq:general_dx_np1}),
and Taylor expand terms involving $\M M,\,\M H,\,\M K$, and $\V h$
evaluated at a point other than $\bar{\V x}^{n}$ to first order.
Note that both options for the handling of the explicit term in (\ref{eq:general_linearizer})
yield the same increment to second order. After performing these expansions
and collecting some terms, we get the final discrete increment to
second order,
\begin{align}
\D{\left(\delta x_{i}\right)}=\d x_{i}^{n+1}-\d x_{i}^{n}= & \D t\, M_{ij}\d x_{j}+\D t^{1/2}K_{ij}\left(\sqrt{w_{2}}W_{j}^{1}+\sqrt{(1-w_{2})}W_{j}^{2}\right)\nonumber \\
 & +\D t^{2}\d x_{j}\left(H_{km}\bar{x}_{m}+h_{k}\right)\times\label{eq:fluctuation_discrete_increment}\\
 & \;\left(\left(w_{2}w_{3}+w_{2}w_{4}\right)\left(\partial_{k}(H_{ij})+\partial_{kj}(H_{il})\bar{x}_{l}\right)+w_{2}w_{6}\partial_{kj}\left(h_{i}\right)+\left(w_{2}w_{3}+w_{4}+w_{5}\right)\partial_{j}(H_{ik})\right)\nonumber \\
 & +\D t^{2}\left(\left(w_{2}w_{3}+w_{4}+w_{5}\right)H_{ij}+\left(w_{2}w_{3}+w_{2}w_{4}\right)\partial_{j}(H_{il})\bar{x}_{l}+w_{2}w_{6}\partial_{j}(h_{i})\right)M_{jk}\d x_{k}\nonumber \\
 & +\D t^{3/2}w_{2}w_{7}\partial_{k}(K_{ij})\left(H_{kl}\bar{x}_{l}+h_{k}\right)\left(\sqrt{w_{2}}\, W_{j}^{1}+\sqrt{1-w_{2}}\, W_{j}^{2}\right)\nonumber \\
 & +\Delta t^{3/2}\left(\left(w_{3}+w_{4}+w_{5}\right)\sqrt{w_{2}}\, W_{k}^{1}+\left(w_{4}+w_{5}\right)\sqrt{1-w_{2}}\, W_{k}^{2}\right)H_{ij}K_{jk}\nonumber \\
 & +\D t^{3/2}\left(w_{3}+w_{4}\right)\sqrt{w_{2}}\,\partial_{k}(H_{ij})\bar{x}_{k}K_{jl}W_{l}^{1}\nonumber \\
 & +\D t^{3/2}w_{6}\sqrt{w_{2}}\partial_{k}(h_{i})K_{kl}W_{l}^{1}+O(\D t^{5/2}),\nonumber 
\end{align}
where we have omitted bars since all quantities and derivatives are
evaluated at $\bar{\V x}^{n}$.

The first moment includes only the deterministic terms, and making
use of the relations (\ref{eq:second_order_cond}) confirms that they
match the continuous increment with error $O(\D t^{3})$. The second
moment is more complicated due to the cross correlations that arise
from the stochastic terms, but after some algebra and again making
use of (\ref{eq:second_order_cond}), we see that the increments match
to the correct order. The third moment only contains the second order
products of the first two terms in the right hand side of (\ref{eq:fluctuation_discrete_increment}).
The fourth moment contains only the fourth moment of the lowest order
stochastic term, $K_{ij}\left(\sqrt{w_{2}\D t}W_{j}^{1}+\sqrt{(1-w_{2})\D t}W_{j}^{2}\right)$.
The fifth moment is zero for both increments. All of these moments
match with error no greater than order $\D t^{3}$, and hence the
scheme (\ref{eq:general_dx_np1}) is second order accurate.

\section{\label{sec:Appendix-Kinetic}Kinetic noise in fast-slow system}

In this section, we show that the schemes (\ref{eq:mid_pred}, \ref{eq:mid_corr})
and (\ref{eq:trap_pred}, \ref{eq:trap_corr}, \ref{eq:trap_stoch_incr_2nd})
from section \ref{sec:Overdamped} produce the correct thermal drift
arising from the kinetic interpretation of the overdamped limit (\ref{eq:limiting_equation}).
It is useful to consider the drift split into the following pieces,
\begin{align}
\left[\partial_{\V x}\cdot\left(\M A+\M B\M C^{-1}\M B^{\star}\right)\right]_{i}=\partial_{j}\left(A_{ij}+B_{ik}C_{kl}^{-1}B_{jl}\right)= & \partial_{j}\left(A_{ik}^{\frac{1}{2}}A_{jk}^{\frac{1}{2}}\right)+\partial_{j}\left(B_{ik}C_{kl}^{-1}\right)B_{jl}+B_{ik}C_{kl}^{-1}\partial_{j}\left(B_{jl}\right),\label{eq:drift_pieces}
\end{align}
where we have rewritten $\M A^{\frac{1}{2}}\equiv\M A_{\frac{1}{2}}$and
$\M C^{\frac{1}{2}}\equiv\M C_{\frac{1}{2}}$ in order to simplify
the notation. The first term on the right hand side of equation (\ref{eq:drift_pieces})
can be split into 
\begin{align}
\partial_{j}\left(A_{ik}^{\frac{1}{2}}A_{jk}^{\frac{1}{2}}\right)= & \partial_{j}\left(A_{ik}^{\frac{1}{2}}\right)A_{jk}^{\frac{1}{2}}+A_{ik}^{\frac{1}{2}}\partial_{j}\left(A_{jk}^{\frac{1}{2}}\right),\label{eq:split_a_drift}
\end{align}
and the second term on the right hand side of equation (\ref{eq:drift_pieces})
can be rewritten as
\begin{align}
\partial_{j}\left(B_{ik}C_{kl}^{-1}\right)B_{jl}= & \partial_{j}\left(B_{ik}C_{kl}^{-1}\right)B_{jn}\delta_{nl}=\partial_{j}\left(B_{ik}C_{kl}^{-1}\right)B_{jn}\left(C_{np}^{-1}C_{pm}^{\frac{1}{2}}C_{lm}^{\frac{1}{2}}\right)\label{eq:split_bcb_drift}\\
= & \partial_{j}\left(B_{ik}C_{kl}^{-1}C_{lm}^{\frac{1}{2}}\right)B_{jn}C_{np}^{-1}C_{pm}^{\frac{1}{2}}-B_{ik}C_{kl}^{-1}\partial_{j}\left(C_{lm}^{\frac{1}{2}}\right)B_{jn}C_{np}^{-1}C_{pm}^{\frac{1}{2}}.\nonumber 
\end{align}
Making use of relations (\ref{eq:split_a_drift}, \ref{eq:split_bcb_drift})
gives the following form of the entire drift 
\begin{align}
\partial_{j}\left(A_{ij}+B_{ik}C_{kl}^{-1}B_{jl}\right)= & \left(\partial_{j}\left(A_{ik}^{\frac{1}{2}}\right)A_{jk}^{\frac{1}{2}}+\partial_{j}\left(B_{ik}C_{kl}^{-1}C_{lm}^{\frac{1}{2}}\right)B_{jn}C_{np}^{-1}C_{pm}^{\frac{1}{2}}\right)\label{eq:split_system_drift}\\
+ & \left(A_{ik}^{\frac{1}{2}}\partial_{j}\left(A_{jk}^{\frac{1}{2}}\right)+B_{ik}C_{kl}^{-1}\partial_{j}\left(B_{jl}\right)-B_{ik}C_{kl}^{-1}\partial_{j}\left(C_{lm}^{\frac{1}{2}}\right)B_{jn}C_{np}^{-1}C_{pm}^{\frac{1}{2}}\right),\nonumber 
\end{align}
 The first line of equation (\ref{eq:split_system_drift}) is generated
by the evaluation of the noise in the corrector step, and the second
line will be approximated using RFD terms. 

Both the implicit trapezoidal and explicit midpoint schemes have the
following noise term in the corrector step,
\begin{align}
x_{i}^{n+1}-x_{i}^{n}= & \ldots+\sqrt{k_{B}T\D t}\left((1-w_{7})A_{ik}^{\frac{1}{2},n}+w_{7}A_{ik}^{\frac{1}{2},p}\right)\left(\sqrt{w_{2}}W_{k}^{n,1,x}+\sqrt{1-w_{2}}W_{k}^{n,2,x}\right)\nonumber \\
 & +\sqrt{k_{B}T\D t}\left((1-w_{7})\left(B_{ik}C_{kl}^{-1}C_{lm}^{\frac{1}{2}}\right)^{n}+w_{7}\left(B_{ik}C_{kl}^{-1}C_{lm}^{\frac{1}{2}}\right)^{p}\right)\left(\sqrt{w_{2}}W_{m}^{n,1,y}+\sqrt{1-w_{2}}W_{m}^{n,2,y}\right),\label{eq:system_noise}
\end{align}
which after Taylor expanding operators evaluated at the predictor
to $O\left(\sqrt{\D t}\right)$ and taking expectation with respect
to $\V W^{x}$ and $\V W^{y}$, becomes
\begin{align}
\avv{x_{i}^{n+1}-x_{i}^{n}}= & \ldots+\left(2w_{2}w_{7}\right)\D t\, k_{B}T\left(\partial_{j}\left(A_{ik}^{\frac{1}{2}}\right)A_{jk}^{\frac{1}{2}}+\partial_{j}\left(B_{ik}C_{kl}^{-1}C_{lm}^{\frac{1}{2}}\right)B_{jn}C_{np}^{-1}C_{pq}^{\frac{1}{2}}\right)+O\left(\D t^{2}\right).\label{eq:expanded_system_noise}
\end{align}
The term on the right-hand side of (\ref{eq:expanded_system_noise})
thus gives the first line of the drift (\ref{eq:split_system_drift})
when $w_{2}w_{7}=1/2$. 

The RFD terms from both schemes give the following additional increment
in $\V x$, 
\begin{align}
x_{i}^{n+1}-x_{i}^{n}= & \ldots+\frac{\D t\, k_{B}T}{\delta}A_{ik}^{\frac{1}{2}}\left(\M A_{jk}^{\frac{1}{2}}(\V x^{n}+\delta\widetilde{\V W})-A_{jk}^{\frac{1}{2}}(\V x^{n})\right)\widetilde{W}_{j}^{A}\nonumber \\
 & +\frac{\D t\, k_{B}T}{\delta}\left(B_{ik}C_{kl}^{-1}\right)^{p}\left(B_{kl}(\V x^{n}+\delta\widetilde{\V W})-B_{kl}(\V x^{n})\right)\widetilde{W}_{m}^{A},\label{eq:system_RFD}\\
 & -\frac{\D t\, k_{B}T}{\delta}\left(B_{ik}C_{kl}^{-1}\right)^{p}\left(C_{lm}^{\frac{1}{2}}\left(\V x^{n}+\delta\M B\M C^{-1}\M C^{\frac{1}{2}}\widetilde{\V W}^{C}\right)\,-C_{lm}^{\frac{1}{2}}(\V x^{n})\right)\widetilde{W}_{m}^{C}
\end{align}
where all terms except for the RFD term have been omitted for this
analysis. Expanding terms not evaluated at $\V x^{n}$ to $O(\delta^{2})$
in (\ref{eq:system_RFD}), and taking expectation with respect to
$\widetilde{\V W}^{A}$ and $\widetilde{\V W}^{C}$ gives
\begin{align*}
\avv{x_{i}^{n+1}-x_{i}^{n}}= & \ldots+\D t\left(k_{B}T\right)\left(A_{ik}^{\frac{1}{2}}\partial_{j}\left(A_{jk}^{\frac{1}{2}}\right)+B_{ik}C_{kl}^{-1}\partial_{j}\left(B_{jl}\right)-B_{ik}C_{kl}^{-1}\partial_{j}\left(C_{lm}^{\frac{1}{2}}\right)B_{jn}C_{np}^{-1}C_{pm}^{\frac{1}{2}}\right),
\end{align*}
which corresponds to the drift term in the second line of equation
(\ref{eq:split_system_drift}). Together, the RFD terms in (\ref{eq:system_RFD})
and the noise terms in (\ref{eq:system_noise}) create the entire
thermal drift from (\ref{eq:limiting_equation}), demonstrating that
our schemes are first order accurate for the fully nonlinear kinetic
equations.


\end{document}